\pdfoutput=1
\documentclass{emulateapj}
\usepackage{natbib}
\shorttitle{HALOGAS:  NGC 4244}
\shortauthors{Zschaechner et al.}

\begin{document}

\title{HALOGAS: {\sc H\,i} Observations and Modeling of the Nearby Edge-on Spiral Galaxy NGC 4244} 
\author{{\sc Laura K. Zschaechner,\altaffilmark{1} Richard J. Rand,\altaffilmark{1} George H. Heald,\altaffilmark{2}\\ Gianfranco Gentile,\altaffilmark{3} and Peter Kamphuis\altaffilmark{4}}}
\altaffiltext{1}{Department of Physics and Astronomy, University of New Mexico, 1919 Lomas Blvd NE, Albuquerque, New Mexico 87131-1156; zschaech@unm.edu, rjr@phys.unm.edu}
\altaffiltext{2}{Netherlands Institute for Radio Astronomy (ASTRON), Postbus 2, 7990 AA Dwingeloo, the Netherlands; heald@astron.nl}
\altaffiltext{3}{Department of Physics and Astronomy, Universiteit Gent, Krijgslaan 281, B--9000 Gent, Belgium; Gianfranco.Gentile@ugent.be}
\altaffiltext{4}{Astronomisches Institut der Ruhr-Universit\"{a}t Bochum, Universit\"{a}tsstr.~150, 44780 Bochum; Germany peter.kamphuis@astro.rub.de}
\slugcomment{Accepted to Astrophysical Journal on July 14, 2011}

\begin{abstract}

\par
     We present 21-cm observations and models of the {\sc H\,i} kinematics and distribution of NGC 4244, a nearby edge-on Scd galaxy observed as part of the Westerbork Hydrogen Accretion in LOcal GAlaxieS (HALOGAS) survey. Our models give insight into the {\sc H\,i} kinematics and distribution with an emphasis on the potential existence of extra-planar gas as well as a negative gradient in rotational velocity with height above the plane of the disk (a lag).  Our models yield strong evidence against a significantly extended halo and instead favor a warp component along the line of sight as an explanation for most of the observed thickening of the disk.  Based on these models, we detect a lag of $-9^{+3}_{-2}$ km s$^{-1}$ kpc$^{-1}$ in the approaching half and $-9\pm$2 km s$^{-1}$ kpc$^{-1}$ in the receding half. This lag decreases in magnitude to $-5\pm$2 km s$^{-1}$ kpc$^{-1}$ and $-4\pm$2 km s$^{-1}$ kpc$^{-1}$ near a radius of 10 kpc in the approaching and receding halves respectively.  Additionally, we detect several distinct morphological and kinematic features including a shell that is probably driven by star formation within the disk.
\end{abstract}


\section{Introduction} \label{Section1}

\par
     Understanding the distribution and kinematics of vertically extended gas in disk galaxies is crucial for comprehending the relevance for the evolution of disk-halo flows (e.g.\ the galactic fountain model presented by \citealt{1976ApJ...205..762S} and refined by \citealt{1980ApJ...236..577B}; and the chimney model of \citealt{1989ApJ...345..372N}), interactions with the intergalactic medium (IGM) such as cold accretion of primordial gas (\citealt{2005MNRAS.363....2K}, \citealt{2008A&ARv..15..189S}) and interactions with neighboring galaxies.  Furthermore, comprehension of the interactions between ISM energization from star formation and accretion, i.e. feedback (e.g.\ \citealt{2000MNRAS.317..697E}) is also imperative.   

\par
    Galactic gaseous halos have been been observed in X-rays (e.g.\ \citealt{2006A&A...448...43T}), radio continuum (e.g.\ \citealt{1999AJ....117.2102I}), dust (e.g.\ \citealt{1999AJ....117.2077H}), ionized hydrogen (e.g.\ \citealt{2003A&A...406..493R}) as well as neutral hydrogen ({\sc H\,i}) (e.g.\ \citealt{1997ApJ...491..140S}, \citealt{2007AJ....134.1019O}).   For the radio continuum (e.g.\ \citealt{2006A&A...457..121D}), diffuse ionized gas (DIG) (e.g.\ \citealt{2003A&A...406..493R}), X-ray (e.g.\ \citealt{2006A&A...457..779T}), and dust \citep{1999AJ....117.2077H} there is a correlation between the presence of these halo components and star formation in the disk, both locally, in regions throughout the galaxy as well as globally.  This favors disk-halo flows akin to those in the aforementioned galactic fountain and chimney models as likely origins.  

\par
    Despite the likely connection between several halo components and star formation, the prevalence and properties of extra-planar {\sc H\,i} in galaxies and any potential connection to star formation is not well understood.  While there are many examples of {\sc H\,i} shells and holes associated with star formation (e.g.\ \citealt{2008A&A...490..555B}), no correlation between the prevalence of widespread {\sc H\,i} halos and star formation activity has been established among multiple galaxies as has been done for other halo components. Clearly, this can only be remedied through greatly expanded observations.

\par
    Another clue to halo origins lies in their kinematics.  Trends concerning extra-planar {\sc H\,i}, such as the presence and magnitude of any decreases in rotation speed with height (lags) or correlations with the kinematics of DIG layers are useful in determining its origins.  One observational issue is whether {\sc H\,i} and DIG halos show the same kinematics, suggesting a common origin.    Recent measurements have shown lags in multiple DIG layers \citep{2007ApJ...663..933H}, leading to various models which attempt to understand this gradient in terms of disk-halo flows and accretion of primordial gas. 

\par
     It has been shown that entirely ballistic models of disk-halo cycling, simulated in \citet{2002ApJ...578...98C} and \citet{2006MNRAS.366..449F}, are too simple in that they are unable to reproduce the observed kinematics.  It is possible the observed motions may partially be due to pressure gradients, magnetic tension \citep{2002ASPC..276..201B} or external influences and feedback such as those described in \citet{2008MNRAS.386..935F}.  Additionally, ``baroclinic" (where gas pressure does not depend on density alone) hydrostatic models considered in \citet{2006A&A...446...61B}, have been able to reproduce the observed lag in NGC 891, as has the hydrodynamic simulation of disk formation by \citet{2006MNRAS.370.1612K}. It is likely that halos form and evolve via a combination of disk-halo flows as well as external influences.

\par
     Of further interest is the radial variation of lags.  Basic geometric arguments predict a shallowing of the lag with increasing radius as may be seen in Figure 3 of \citet{2002ApJ...578...98C}, but this could be complicated, for example, by pressure gradients \citep{2002ASPC..276..201B}.  This will be discussed in greater detail in $\S$~\ref{Section6.2}.

\par
     Detailed kinematic modeling of deep observations of {\sc H\,i} emission has been done for only a small number of galaxies including NGC 891 (\citealt{1997ApJ...491..140S}, \citealt{2007AJ....134.1019O}), NGC 5746 \citep{2008ApJ...676..991R}, and NGC 2403 (\citealt{2002AJ....123.3124F}). Through the Westerbork Hydrogen Accretion in LOcal GAlaxies (HALOGAS) survey \citep{2011A&A...526A.118H},  which targets 22 edge-on or moderately inclined spiral galaxies for deep 10$\times$12 hour observations using the Westerbork Synthesis Radio Telescope (WSRT), we aim to substantially increase this number in order to ascertain information concerning the origins of neutral extra-planar gas.  Through deep observations and kinematic modeling of these galaxies, we will establish whether there is any connection between extra-planar gas and star formation as well as the degree to which contributions from external origins are relevant.  

\par
\begin{figure*}
\epsscale{1}
\figurenum{1}
\includegraphics[width=170mm]{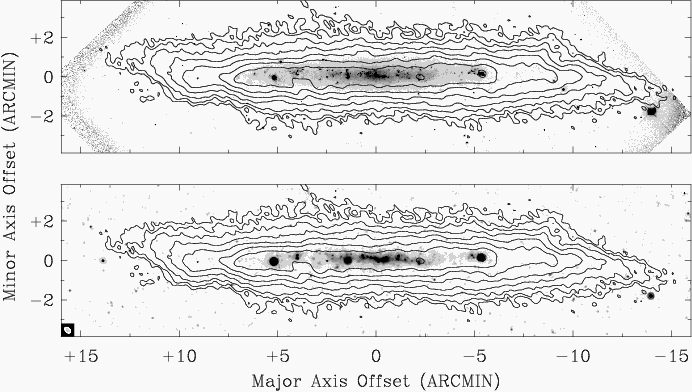}
\caption{The {\sc H\,i} zeroth-moment map overlaid on H$\alpha$ (top) and MIPS 24 $\mu$m (bottom) images.  Each image is rotated counter-clockwise by 43$\,^{\circ}$. {\sc H\,i} contours begin at $6.4\times10^{19}\mathrm{cm^{-2}}$ and increase by factors of 2.  The {\sc H\,i} beam is shown in the lower left-hand corner in white. The H$\alpha$ image is supplied by Rene Walterbos from \citet{1999ApJ...522..669H}. \label{fig1}}
\end{figure*}

\begin{figure}
\begin{center}
\figurenum{2}
\includegraphics[width=80mm]{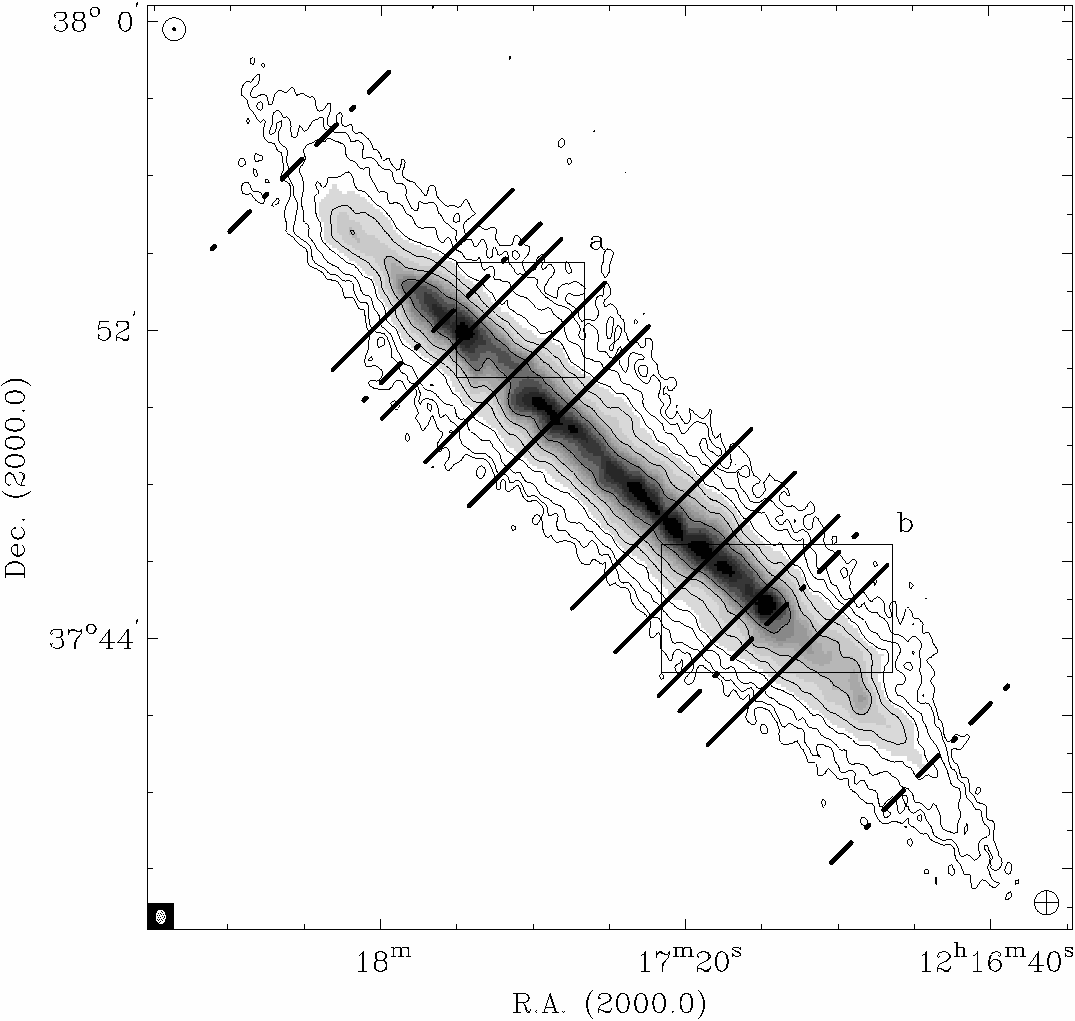}
\caption{A zeroth-moment map of NGC 4244 showing the slice locations for the bv plots in Figure~\ref{fig6} (solid), as well as the range of the evenly spaced slices in Figure~\ref{fig10} (dashed).  Boxes a and b correspond to the regions shown in Figures~\ref{fig11} and~\ref{fig13} respectively. Contours begin at $6.4\times10^{19}\mathrm{cm^{-2}}$ and increase by factors of 2. The {\sc H\,i} beam is shown in the lower left-hand corner in white. An encircled dot and cross denote the approaching and receding halves respectively.\label{fig2}}
\end{center}
\end{figure}

  
     As part of the HALOGAS survey, we present observations and models of NGC 4244, a nearby edge-on Scd galaxy, which exhibits a thickened {\sc H\,i} layer upon inspection of the zeroth-moment map (presented here, Figures~\ref{fig1},~\ref{fig2}) as well as in previous work \citep{1996AJ....112..457O}.  The first goal of the modeling is to determine the cause of the observed thickening, whether it is due to the presence of an {\sc H\,i} halo, flare, warp along the line of sight, or any combination of these possibilities. Second, we constrain whether a lag exists. Finally, we will identify local features, such as possible energy injection sites, which will be discussed further in $\S$~\ref{Section6.4}. This work will aid in establishing trends concerning the presence of halos and lags and any connection between disk-halo interactions and/or accretion.  In the case of NGC 4244, which has a low star formation rate (0.12 M$_{\odot}$ yr$^{-1}$; \citealt{2011A&A...526A.118H}), we are especially concerned with the potential existence of a lag.  If the trend found by \citet{2007ApJ...663..933H} for DIG layers in galaxies with low star formation rate resulting in steeper lags holds true for {\sc H\,i}, then a steep lag is expected.

\section{Observations and Data Reduction} \label{Section2}

\par
     Here we provide a brief explanation of the observations and data reduction process.  A detailed description, which applies to all of the HALOGAS galaxies including NGC 4244, may be found in \citet{2011A&A...526A.118H}.      

\par
     Observations of NGC 4244 were primarily obtained as part of the ongoing HALOGAS survey. To best observe faint extended emission, the array was used in Maxishort configuration with baselines ranging from 36 m to 2.7 km for $6\times12$ hours.  Of the 10 fixed antennas, 9 were used and were spaced at 144 m intervals on a regular grid.  Additionally, $4\times12$ hours of archival data \citep{2005A&A...432..475D}, observed in the traditional Westerbork variable configuration are used. The total bandwidth is 10 MHz with 1024 channels and two linear polarizations.  

\par
  Data reduction was done in Miriad \citep{1995ASPC...77..433S}. The {\sc H\,i} data were imaged with a variety of weighting schemes, creating multiple data cubes.  Clark deconvolution \citep{1980A&A....89..377C} was performed using mask regions defined on the basis of unsmoothed versions of the data.  Offline Hanning smoothing led to a final velocity resolution of 4.12 km s$^{-1}$.  The 1$\sigma$ rms noise in a single channel of the full resolution cube is 0.22 mJy bm$^{-1}$, or corresponding to a column density of  $N_{HI}=3.5\times10^{18}$ cm$^{-2}$.

\par
    For the primary modeling, we use a cube with a robust parameter of 0 for intermediate resolution and sensitivity, with a 21"$\times$13.5" beam, having a position angle of $-0.7\,^{\circ}$. Additionally, to maximize sensitivity to faint extended emission we use a Gaussian {\it uv} taper corresponding to 30" resolution in the image plane to create a second cube.  Further smoothing to 60" resolution yielded negligible additional extended emission.  Arcseconds are converted into parsecs via multiplying by 21.35 pc arcsecond$^{-1}$ [obtained by assuming a distance of 4.4 Mpc (Table~\ref{tbl-1})], resulting in a resolution of approximately 450$\times$290 pc for the full resolution cube.  

\begin{deluxetable*}{lrr}
\tabletypesize{\scriptsize}
\tablecaption{NGC 4244 Parameters\label{tbl-1}}
\tablewidth{0pt}
\tablehead
{
\colhead{Parameter} &
\colhead{Value}&
\colhead{Reference}\\
}
\startdata
\phd Distance (Mpc) &4.4\tablenotemark{a} &Heald et al. (2011)\\ 
\phd Systemic velocity (km s$^{-1}$)&244 &\citet{1996AJ....112..457O}, This work\\
\phd Inclination &88$\,^{\circ}$ &This work\\
\phd SFR (M$_{\odot}$ yr$^{-1}$) &0.12 &Heald et al. (2011)\\ 
\phd Morphological Type &Scd & \citet{1991trcb.book.....D} \\
\phd Kinematic Center $\alpha$ (J2000.0) & 12h17m29.90s &This work\\
\phd Kinematic Center $\delta$ (J2000.0) &37d48m29.00s &This work\\
\phd Optical Radius (kpc) &10.4\tablenotemark{b} &\citet{1999AJ....118.1209F}\\
\phd Total Atomic Gas Mass ($10^{9}M_{\odot}$) & 2.5 &This work\tablenotemark{c}\\
\phd Total Atomic Gas Mass  ($10^{9}M_{\odot}$) & 2.2 &\citet{1996AJ....112..457O}\tablenotemark{c}\tablenotemark{d}\\
\enddata
\tablenotetext{a}{Distance is the median value of distances found on the NED database, excluding those obtained using the Tully-Fisher relation.}
\tablenotetext{b}{This value is adjusted for the difference in distance of 4.4 Mpc used in this paper and 3.6 Mpc used in \citet{1999AJ....118.1209F}}
\tablenotetext{c}{Includes neutral He via a multiplying factor of 1.36.}
\tablenotetext{d}{Scaled to our assumed distance.}
\end{deluxetable*}

\par
A summary of observational and data reduction parameters may be found in Table~\ref{tbl-2}.

\begin{deluxetable}{lr}
\tabletypesize{\scriptsize}
\tablecaption{Observational and Instrumental Parameters  \label{tbl-2}}
\tablewidth{0pt}
\tablehead
{
\colhead{Parameter} &
\colhead{Value}\\
}
\startdata
\phd Observation Dates $-$ HALOGAS&2009 Jul 13\\
\phd &2009 Jul 16\\
\phd &2009 Jul 18/Aug 1\\
\phd &2009 Nov 15\\
\phd &2009 Nov 27\\
\phd Observation Dates $-$ Dahlem&2003 Feb 13-14\\
\phd &2003 Feb 19-20\\
\phd &2003 Feb 20-21\\
\phd &2003 Apr 13-14\\
\phd Pointing Center &12h17m29.901s\\
\phd  &37d48m29.00s\\ 
\phd Number of channels &68\\
\phd Velocity Resolution &4.12 km s$^{-1}$\\
\phd Full Resolution Beam Size &21$\times$13.5"\\
\phd &450$\times$290 pc\\
\phd RMS Noise $-$ 1 Channel (21$\times$13.5")&0.22 mJy bm$^{-1}$\\
\phd RMS Noise $-$ 1 Channel (30$\times$30")&0.37 mJy bm$^{-1}$\\
\enddata
\end{deluxetable}

\section{The Data} \label{Section3}

\par
     The full resolution zeroth-moment map is produced using the Gronigen Image Processing System (GIPSY; \citealt{1992ASPC...25..131V} task {\tt moments} after applying a mask to the full resolution cube based on the 3$\sigma$ limit of a 90" smoothed cube.  This is shown overlaid onto H$\alpha$ \citep{1999ApJ...522..669H} and 24 $\mu$m MIPS (NASA/IPAC Infrared Science Archive) images in Figure~\ref{fig1} as well as in grayscale in Figure~\ref{fig2}.  A warp component that is perpendicular to the line of sight is apparent.  Also, as will be shown through modeling described in $\S$~\ref{Section4}, the minor axis extent is largely due to a warp component along the line of sight.   Minor asymmetries may be noted, such as slight differences in the warped disk in the approaching (northeast) and receding (southwest) halves as well as more radially extended emission in the receding half.  Additional localized asymmetries and irregularities are also present throughout.  The 30" map (not shown) displays the same indications of a warp as well as the previously mentioned asymmetries.  Even with the increased sensitivity of this map, it does not reveal any substantial faint emission not already seen in the full resolution cube.
     
\par
\begin{figure*}
\begin{center}
\figurenum{3}
\includegraphics[width=170mm]{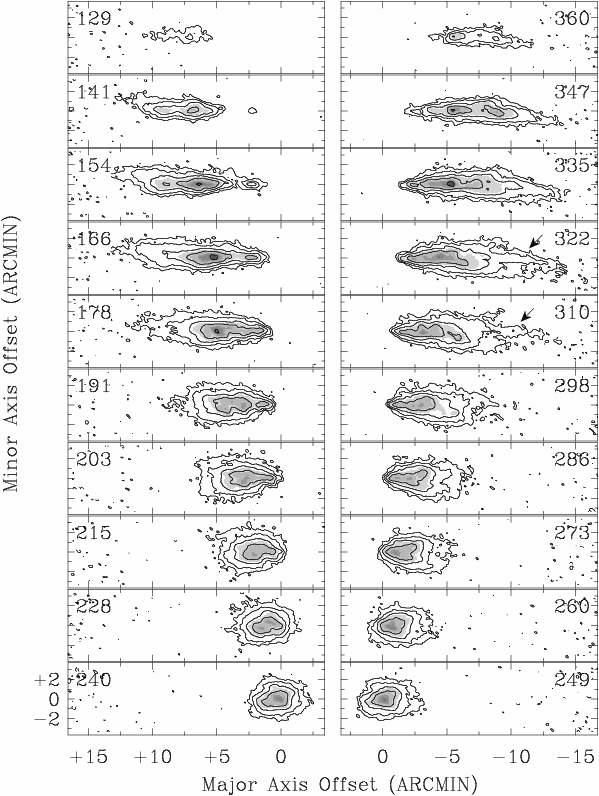}
\caption{Representative channel maps of the data with a resolution of 21"$\times$13.5".  Contours begin at 3$\sigma$ (0.66 mJy bm$^{-1}$) and increase by factors of 3. Negative contours are also shown with dashed lines. The velocities in km s$^{-1}$ are given in each panel and $\delta{v}$ is approximately 3 resolution elements.  Note the slanted outer edges present in both halves, which correspond to the projection of the warp perpendicular to the line of sight.  Arrows indicate the pronged streak discussed in the text in the lowest contours of the channels corresponding to 310 and 322 km s$^{-1}$. \label{fig3}}
\end{center}
\end{figure*}
\begin{figure}
\begin{center}
\figurenum{4}
\includegraphics[scale=.55]{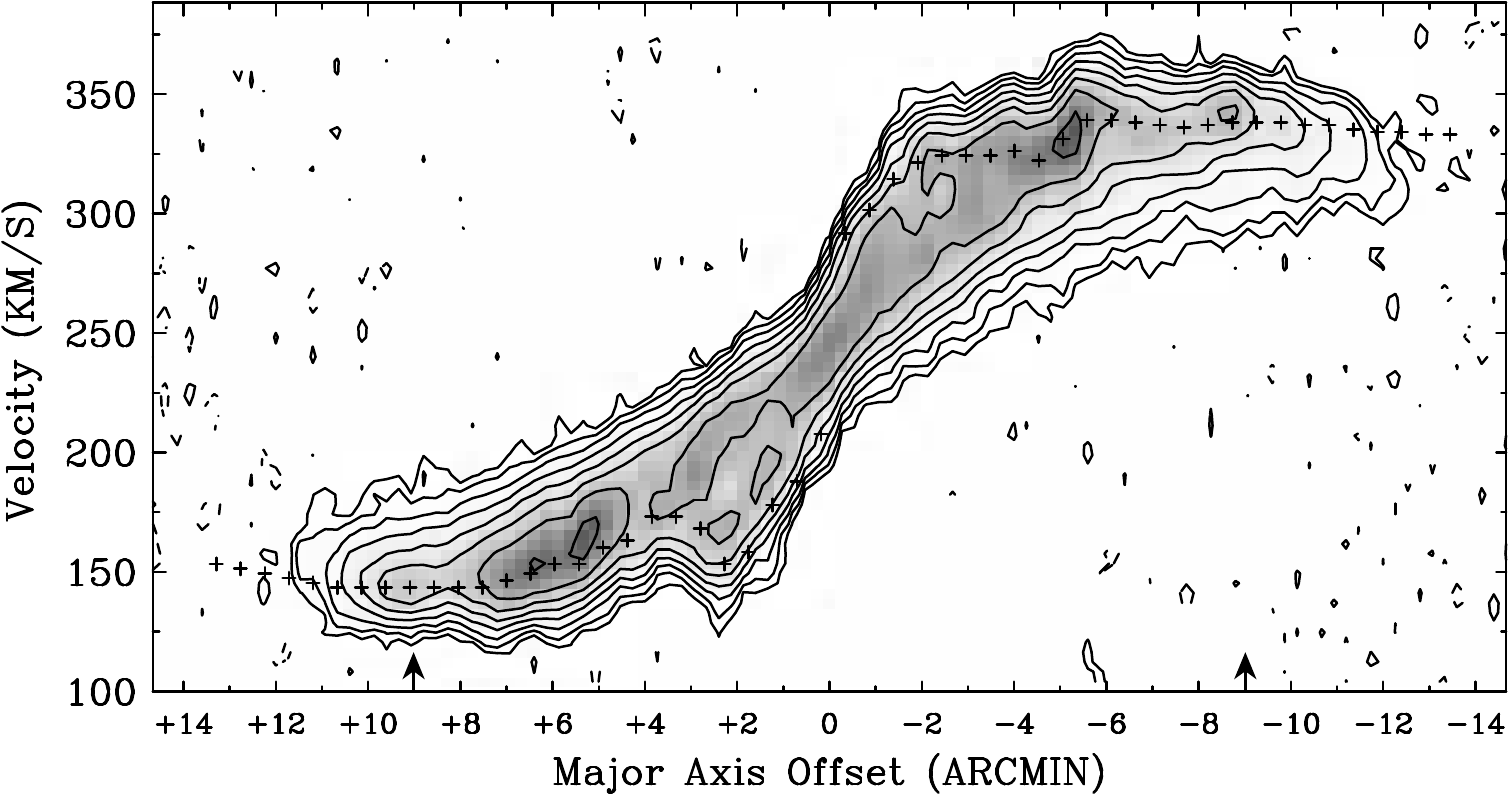}
\label{fig:first_sub}
\caption{Major axis position-velocity diagram.  Contours begin at 2$\sigma$ (0.44 mJy bm$^{-1}$) and increase by factors of 2.  Asymmetries are once again prevalent in the two halves. The rotation curve for each half is shown with crosses.  Arrows near the horizontal axis indicate the approximate starting point of the radial decrease in the lag discussed in $\S$~\ref{Section4.2.6}. The warp component across the line of sight creates the appearance that the rotation curve extends beyond the data, but at heights above the midplane, these points are constrained.  \label{fig4}}
\end{center}
\end{figure}

     Full resolution representative channel maps are provided in Figure~\ref{fig3}.  Note the faint streak-like emission in the lowest contours extending radially outward from the center of the galaxy in the 310 km s$^{-1}$ and 322 km s$^{-1}$ channels.  Figure~\ref{fig4} shows a position-velocity diagram along the major axis (lv diagram) with the rotation curve derived through modeling. The previously mentioned asymmetries are also easily seen in the channel maps as well as the lv diagram.

\par
     Careful inspection of the data lays the foundation for the modeling process described below.

\section{Models} \label{Section4}

\par
    To extract information from the data useful in constraining the {\sc H\,i} distribution and kinematics, including a possible lag, model data cubes are created using GIPSY.  Using the task {\tt galmod}, a model galaxy is divided into concentric rings where the column density, rotational velocity, scale height, velocity dispersion, inclination and position angle for each are specified.  Additionally, we use a version of {\tt galmod}, modified by one of us (G. H.) to allow a linear gradient in rotational velocity with height. These tasks assume axisymmetry, which is only approximately true for NGC 4244 (best seen in Figures~\ref{fig1} through~\ref{fig4}). Therefore, the approaching and receding halves are modeled separately, which greatly improves the models for each half but does not completely eliminate issues due to asymmetries as will be demonstrated. These remaining issues however, do not impact our final result concerning the existence and magnitude of a lag.
     
\par
     For initial estimates of the column densities as well as the rotation curve, we used the models of \citet{1996AJ....112..457O}, who found evidence for a flaring gas layer.  Given the improved depth of our data by comparison (with the rms noise for this data being 0.22 mJy bm$^{-1}$ compared to 1.9 mJy bm$^{-1}$ for a cube of similar resolution (28.3$\times$10.1") used by Olling) as well as differences in modeling techniques used, it is possible for the results to differ.  

\par
\begin{figure}
\begin{center}
\figurenum{5}         
\includegraphics[scale=.55]{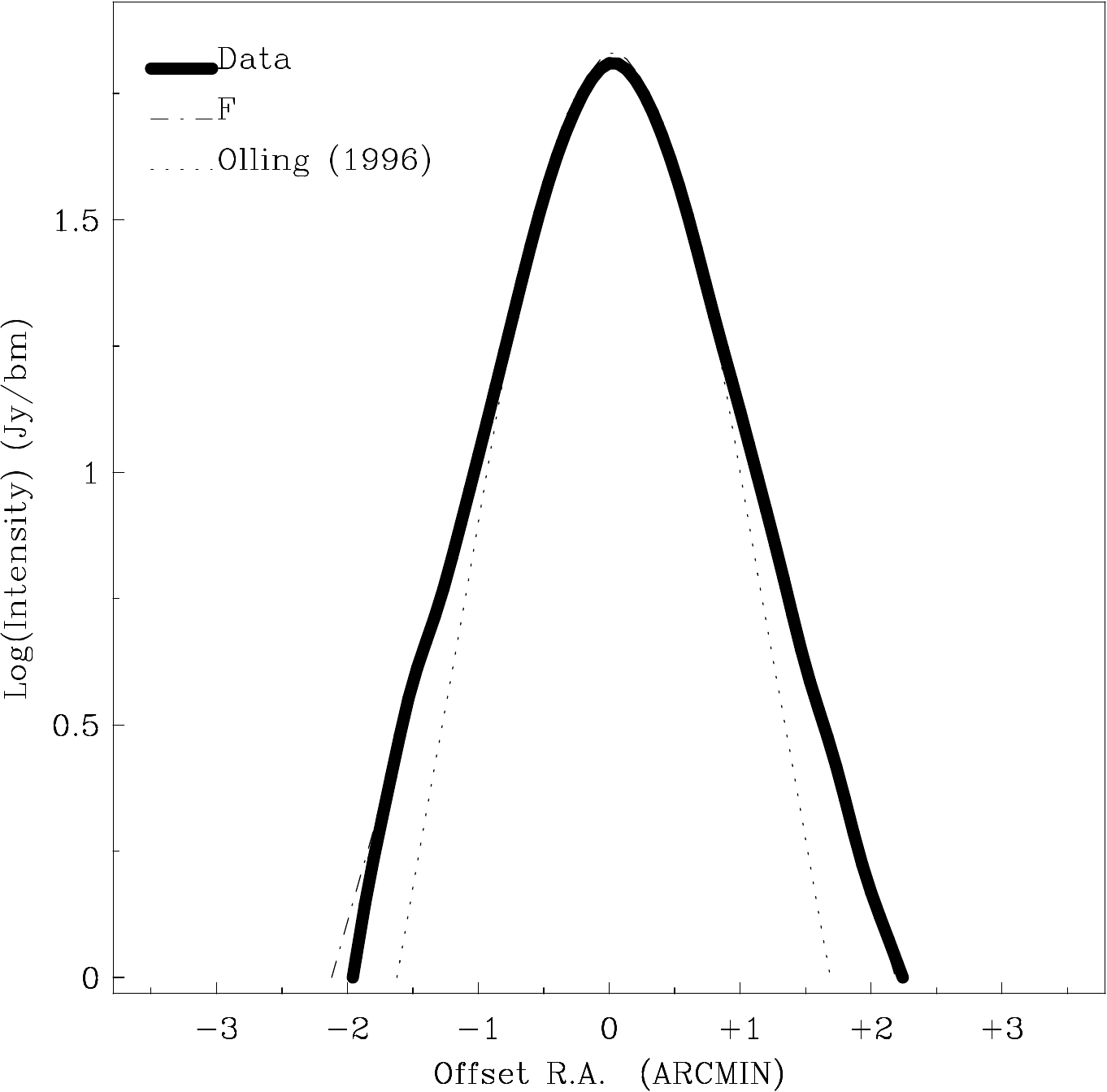}    
  \caption{Log-scale vertical profiles of the data, flaring model and Olling's 1996 model with a constant inclination for both halves.  The other models in this paper are not shown as they match the data closely, to within one line width.  Notice how the data are fit with a single exponential and do not display prominent emission in the wings. \label{fig5}}
\end{center}
\end{figure}

     Rough initial estimates for the kinematic center and systemic velocity were found using the zeroth and first moment maps.  Estimates for the position angles were determined via inspection of the zeroth-moment map and were immediately refined prior to altering additional parameters.  The estimated column densities are refined first by matching both the shape and total flux in the vertical profile (Figure~\ref{fig5}, made by including emission only within a major-axis distance of 5' from the center to avoid effects such as the warp and tapering of the disk found at larger radii) and zeroth-moment maps of the model with those of the data.

\par
     Provided a good fit for previous steps, the rotation curve is then honed primarily through the use of lv diagrams, tracing emission along the edge with the highest velocities (terminal side).  These, along with minor axis (bv) position-velocity diagrams, as well as channel maps of models and data must all be examined to fine tune the above parameters in the models, while at the same time ensuring that the vertical profiles and zeroth-moment maps are well matched.  Although all plots are considered throughout the process, the bv diagrams and channel maps are the most heavily relied upon after the initial modeling stages. 

\par
     As will be demonstrated, the key to this method is to start with the simplest possible model and then add features such as warps and flares - one at a time - to fully understand the individual contribution of each additional degree of complexity. This allows for subtle distinctions between features, which otherwise may be misinterpreted. In the end, an acceptable model is one that fits \textit{all} of the plots well.  The quoted errors indicate visually estimated uncertainties rather than formal 1$\sigma$ error bars.

\par
    It should be noted that new, semi-automated tilted ring fitting software ({\tt TiRiFiC}) \citep{2007A&A...468..731J}, which allows for a chi-square fit of models to observed data cubes is now available.  Advanced models for NGC 4244 were examined in both {\tt TiRiFiC} and {\tt galmod} and each produced comparable results.  Given the advanced stage of modeling already completed via {\tt galmod}, as well as the localized features in NGC 4244, {\tt TiRiFiC} was not used extensively in this case.  However, the advent of {\tt TiRiFiC}'s automated processes will help to expedite the fitting of the remaining galaxies in the HALOGAS sample.  

\subsection{Features Considered While Modeling} \label{Section4.1}   
\par
\begin{figure*}
\begin{center}
\figurenum{6}
\includegraphics[angle=270, width=170mm]{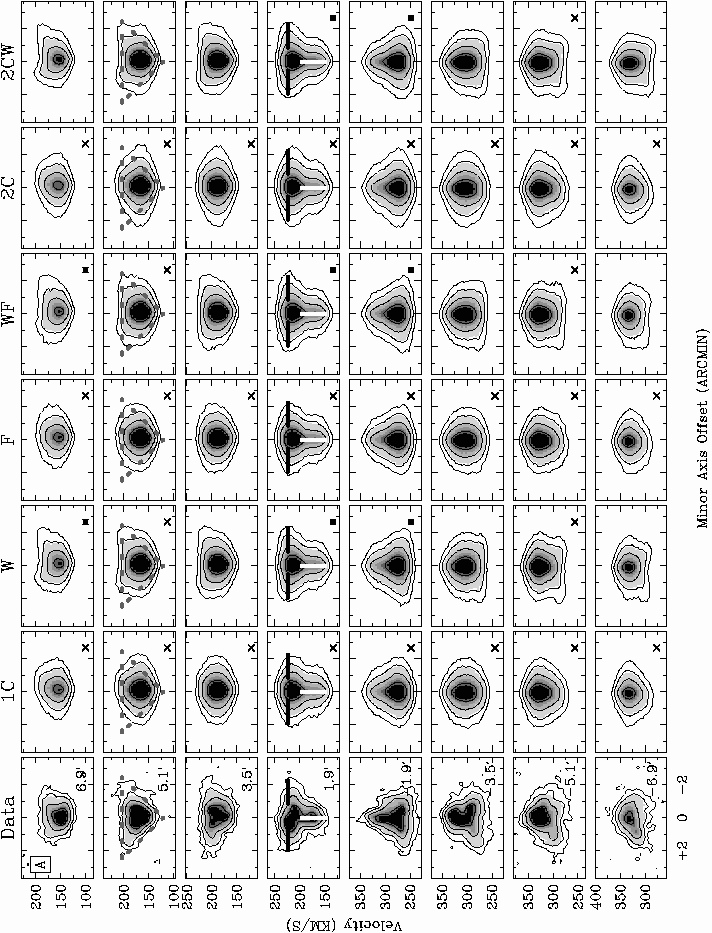}
\caption{Minor axis position-velocity diagrams of NGC 4244 and models.  Slice locations are those displayed by solid lines in Figure~\ref{fig2} and velocity ranges vary from panel to panel.  Contours begin at 3$\sigma$ (0.66 mJy bm$^{-1}$) and increase by factors of 3.  All models include a projection of a warp across the line of sight.  The W model is favored and others are shown for illustrative purposes only.  The symbols provided in the lower right-hand corners of selected panels indicate quality of fit to the data judged by eye.  Crosses indicate a very poor fit, blank a somewhat poor fit, boxes a reasonable one, and asterisks a good fit.  The ``V" shape and flattened systemic side are emphasized with gray dotted lines, while the ``T" shape is shown in black and white lines.  (A) displays models with no lags, while (B) shows analogous plots, but with the addition of lags as well as the omission of the 2C model.  The magnitudes of lags are $-$7 km s$^{-1}$ kpc$^{-1}$ for both halves of 1CL as well as FL, $-$9 km s$^{-1}$ kpc$^{-1}$ for the WL and WFL models, and $-$9 km s$^{-1}$ kpc$^{-1}$ for the 2CWL model.  Due to local asymmetries, the emission extending on the negative minor axis offset near 125 km s$^{-1}$ in the panel corresponding to 6.9' is not considered while modeling.  Additionally, on the receding half, the side with a positive offset is given less weight in the fitting process for the same reason.\label{fig6}}
\end{center}
\end{figure*}

     The bv diagrams shown in Figure~\ref{fig6} and the channel maps in Figure~\ref{fig7} illustrate a number of the features that were noticeably sensitive to the model parameters we endeavor to match.  There is a notable ``T" shape in the panels closest to the galactic center in the bv plots in Figure~\ref{fig6}, which morphs into a ``V" shape in the outer panels (schematically indicated in certain panels of Figure~\ref{fig6}).  Additionally, the contours on the systemic side are flattened.  There is also a substantial lopsidedness best seen at $-$5.1' and $-$6.9' in Figure~\ref{fig6} around the midplane, primarily due to the component of the warp perpendicular to the line of sight, but also partially due to asymmetries. 

\begin{figure*}
\begin{center}
\includegraphics[angle=270,width=145mm]{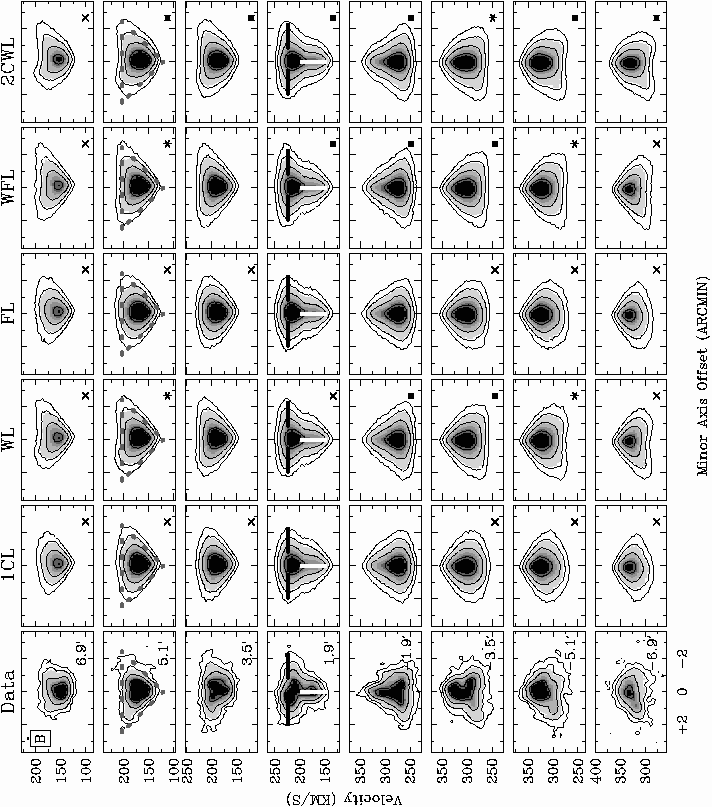}
\label{fig6b}
\end{center}
\end{figure*}

\begin{figure*}
\begin{center}
\figurenum{7}
\includegraphics[angle=270, width=170mm]{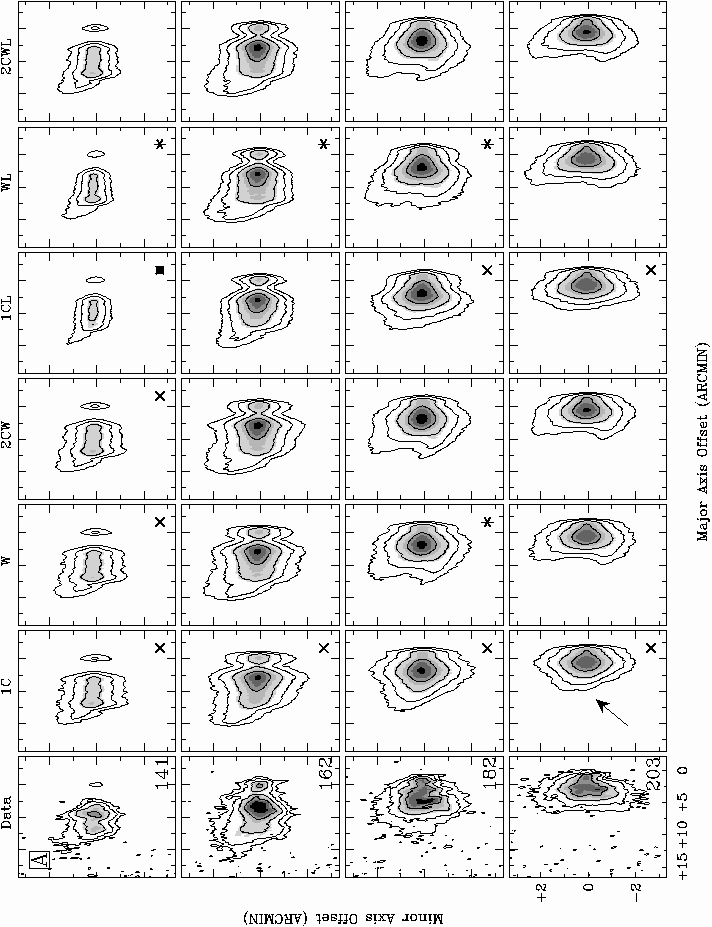}
\caption{Channel maps of NGC 4244 compared with models.  Contours begin at 3$\sigma$ (0.66 mJy bm$^{-1}$) and increase by factors of 3. Velocities are given in the first column in kilometers per second. All models include a projection of a warp perpendicular to the line of sight. Lag models are denoted with an L.  (See Figure~\ref{fig6} caption for the lag magnitudes.) The pure flare and pure two-component models are not included as they are shown to be poor matches in the bv plots.  Finally, the WF model has also been excluded as it is similar to the warp only model.  The definition of symbols follows that set forth in the caption of Figure~\ref{fig6}.  The approaching half is shown in (A) while the receding half is shown in (B). \label{fig7}}
\end{center}
\end{figure*}

\begin{figure*}
\begin{center}
\includegraphics[angle=270, width=170mm]{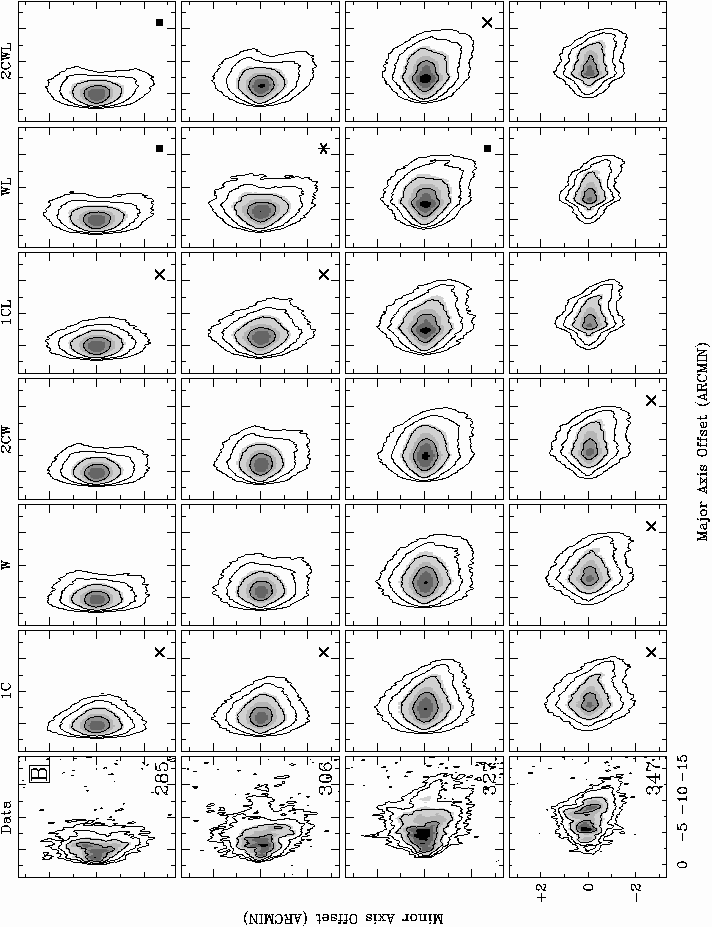}
\label{fig7b}
\end{center}
\end{figure*}
     
\par
     The channel maps displayed in Figure~\ref{fig7} reiterate the need for a warp perpendicular to the line of sight. Prominent features that were sensitive to model parameters considered during the modeling process include the spacing and angle of the contours at the tips of each plot closest to the center of the galaxy.  Also considered is the approximately 45 degree slant along the outermost edges at more extreme velocities and at mostly negative minor axis offsets in the approaching half and positive offsets in the receding half.

\par  
     In both Figures~\ref{fig6} and~\ref{fig7}, the visual assessment of the quality of the fit between the models and data is indicated in each panel via symbols explained in the caption of Figure~\ref{fig6}.  The assignment of each symbol heavily relies on the criteria described above.  

\par
\begin{figure}
\begin{center}
\figurenum{8}
\includegraphics[scale=.55]{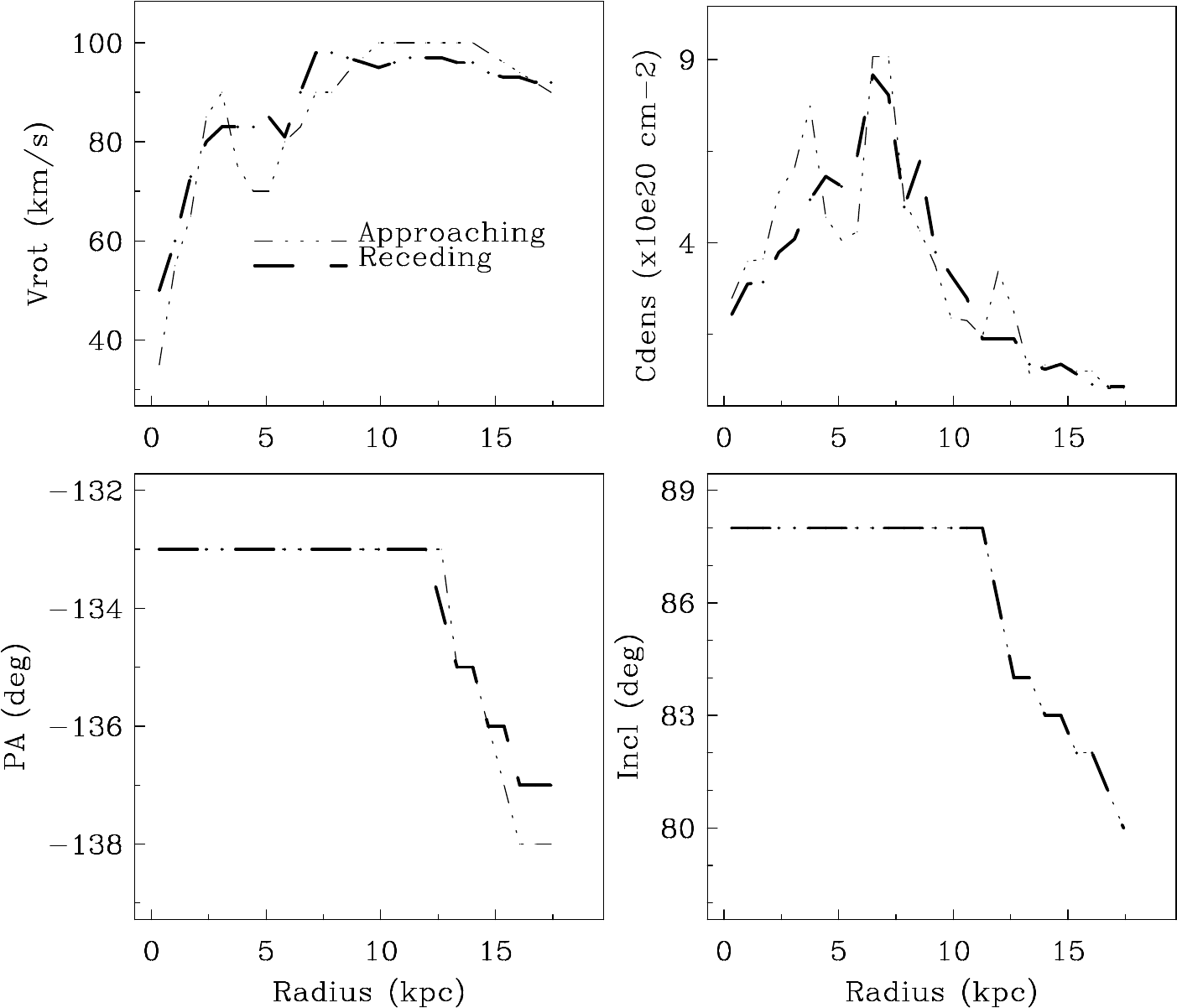}          
\caption{Shown are parameters used for the models.  The rotation curves, column densities and position angles presented are consistent for all models.  The inclinations are those used for the warp and two-component models.}
  \label{fig8}
\end{center}
\end{figure}

     Some quantities are kept constant throughout all of the models for the entire process.  These are the systemic velocity (244 km s$^{-1}$), kinematic center (12h17m29.90s, 37d48m29.00s) and the run of the position angle derived from the component of the warp perpendicular to the line of sight, column density profile, rotation curve (both shown in Figure~\ref{fig8}) and velocity dispersion (12 km s$^{-1}$ decreasing to 10 km s$^{-1}$ at a radius of ~7.7' or 10 kpc). 

\subsection {Individual Models} \label{Section4.2}     

\par
     Several possible models are considered. Some are eliminated quickly upon inspection of the vertical profile or zeroth-moment map, while others require substantially more effort to discern between them.  Vertical distributions are exponentials in all models.  Exponential scale heights vary among models and are listed in Table~\ref{tbl-3}. All of the models, unless otherwise noted, match the zeroth-moment map and vertical profile reasonably well, as will be graphically shown in Figure~\ref{fig9} and in the following discussion.

\begin{deluxetable}{lr}
\tabletypesize{\scriptsize}
\tablecaption{Model exponential scale heights rounded to the nearest 25 pc.\label{tbl-3}}
\tablewidth{0pt}
\tablehead
{
\colhead{Parameter} &
\colhead{Value}\\
}
\startdata
\phd scale height $-$ 1C (pc) &600\\ 
\phd scale height $-$ F (pc) &550 $\rightarrow$ 1100\tablenotemark{a}\\
\phd scale height $-$ W (pc) &550, 575\tablenotemark{b}\\
\phd scale height $-$ WF (pc) &550 $\rightarrow$ 1100\tablenotemark{a}\\
\phd scale height $-$ 2C (pc) & 525, 1050\tablenotemark{c}\\
\phd scale height $-$ 2CW (pc) &425, 850\tablenotemark{c}\\
\enddata
\tablenotetext{a}{the first value applies to radii within the optical radius.  Beyond this radius, the scale height linearly increases until the final value (twice the initial) is reached.}
\tablenotetext{b}{Values for the approaching and receding halves respectively.}
\tablenotetext{c}{Values for the thin and thick components respectively.  These are consistent for both halves.}
\end{deluxetable}

\subsubsection{One-component Model} \label{Section4.2.1}

\par
\begin{figure}
\figurenum{9}
\includegraphics[width=80mm]{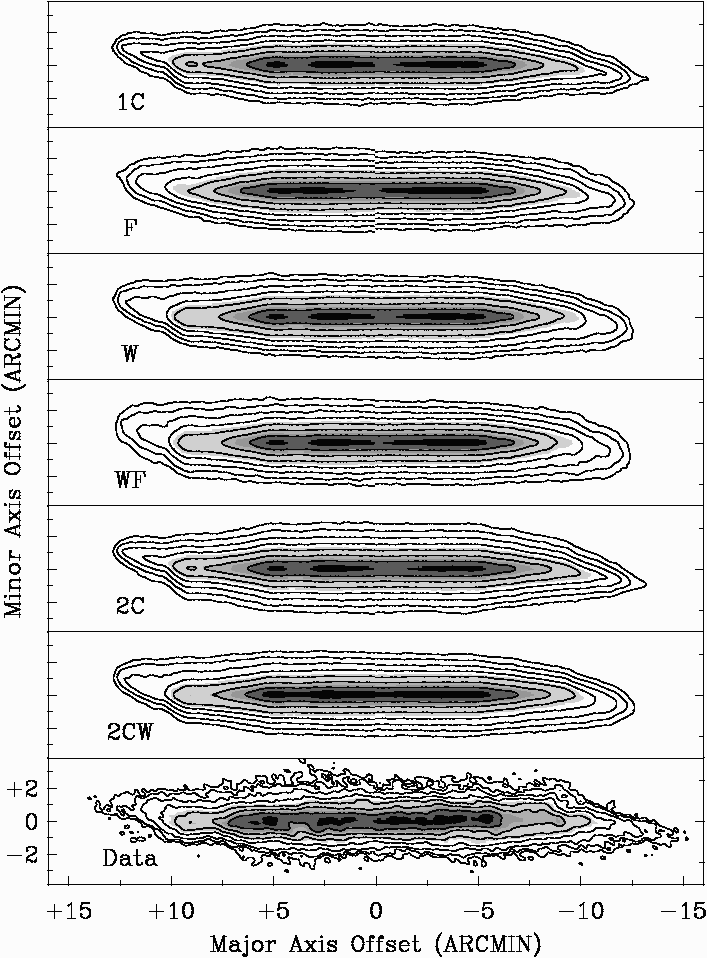}
\caption{Zeroth-moment maps of the data and models.  The model plots are comprised of both the approaching (positive offset) and receding (negative) separately modeled halves. Contours begin at $6.4\times10^{19}\mathrm{cm^{-2}}$ and increase by factors of 2.  \label{fig9}}
\end{figure}

   The simplest model involves a single disk with a slight warp perpendicular to the line of sight [deviating by 4-5$\,^{\circ}$ from the inner disk (Figure ~\ref{fig8}) and beginning at a radius of 12 kpc].  For such a model, all other quantities aside from column densities and rotational velocities are held constant with radius. The inclination is found to be $88\,^{\circ}$ for the best fit to the vertical profile.  For the final model, the scale height is about 600 pc.  This simple model (referred to as ``1C" in figures) cannot be immediately eliminated based on the fit to the vertical profile and zeroth-moment map (Figure~\ref{fig9}).  However, one may note that in the bv plots in Figure~\ref{fig6}, the one-component model lacks sufficient flux on the systemic relative to the terminal side in each panel.  As can be seen in the channel maps in Figure~\ref{fig7}, shown most prominently in the plots closest to the systemic velocity and noted with an arrow, the edge of the diagram away from the center of the galaxy comes to a definitive point at a positive minor axis offset on the approaching half, and a negative offset on the receding, which is not seen in the data.   No other combination of inclination and scale height improves this.  

\subsubsection{Warp Along the Line of Sight} \label{Section4.2.2}

\par  
     Adding a projection of the warp along the line of sight (referred to as ``W" in figures), by gradually decreasing the inclination in the outermost rings (shown in Figure~\ref{fig8}), matches the observed ``T" shape from the data in the central bv diagrams in Figure~\ref{fig6} significantly better than the 1C model.  This is the result of bringing flux from outer rings with lower projected velocity into higher latitude regions.  If the inclination decreases too quickly, the ``T" shape becomes excessive. 
        
\par
     In contrast to the one-component model, the channel maps in Figure~\ref{fig7} show the flux is pushed upward and at high z pushed away from the center in the major axis direction, in panels with velocities close to the systemic velocity in the warp model, which matches the data more closely. However, if too much of a warp is present in the model, then analogous to the ``T" shape verging on becoming too ``Y"-shaped in the bv diagrams, there will be an indentation on this edge.  This is not seen in the data. 

\par
    The best fit LOS warp component deviates by 8 degrees from the inner disk and as in the PA warp, starts at a radius of 12 kpc (Figure~\ref{fig8}).
     
\subsubsection{Flaring Model} \label{Section4.2.3}

\par
     Upon initial inspection of the zeroth-moment maps, it may be noted that the edges in the outer radii taper significantly in minor axis extent.  Because of this, a substantial flare can be ruled out immediately as it would cause the zeroth-moment map to become too boxy and fail to fit this aspect of the data.  However, this is not to say for certain that a modest flare may not be present.  For this reason, a representative flaring model (referred to as ``F" in figures) is created by starting with the one-component model and increasing the scale height outward radially.  This increase begins at the optical radius (10.4 kpc; Table~\ref{tbl-1}) and increases linearly, approximately doubling by a radius of 15 kpc.  Comparison of zeroth-moment maps from the data (Figure~\ref{fig9}) and model indicates that a modest flare yields a slight improvement compared to the one-component model.  Unlike the warp model, the modest flare does not eliminate the rounding on the systemic side of the bv plots (Figure~\ref{fig6}) or the thinning on the outer edges of the channel maps seen in the one-component model.  A more extreme and less physical flare (abruptly doubling the scale height at the optical radius and then holding it constant at larger radii) helps to reduce these problems in the bv plots and channel maps, but once again, causes the zeroth-moment map to become boxy and an obvious mismatch to the data (not shown).  

\par 
     The pure flare model clearly does not match the data in the bv plots.  However, adding the same modest flare to the warp model (referred to as ``WF" in figures) does not appear to have a substantial effect; in fact, it is almost identical to the warp model as shown in the zeroth-moment map in Figure~\ref{fig9} as well as the bv plots in Figure~\ref{fig6}.  This indicates that such a flare could be present in NGC 4244, but that the component of the warp along the line of sight is dominant and effectively hides most of its contribution.  Since we are unable to determine whether or not the flare exists due to the minimal contribution it would have compared to the warp, and for the sake of clarity, it is omitted from further models.

\subsubsection{Two-component Model} \label{Section4.2.4}

\par
     {\sc H\,i} halos have been observed in several galaxies, with NGC 891 among the most notable \citep{2007AJ....134.1019O}.  One of the primary goals of HALOGAS is to determine the frequency of occurrence and origins of such halos.  We therefore examine the possibility of such a halo in NGC 4244. 

\par
     The addition of a halo to the model is made via combining a second, thicker exponential component to the first, keeping the inclination of both at $88\,^{\circ}$ (referred to as ``2C" in figures). Upon inspection of the vertical profile and zeroth-moment map, there is no immediate motivation for a second component, as reasonable fits can be made with less complicated models. For both halves of NGC 4244,  multiple ratios of scale height and flux in each component were tested in attempts to match the vertical profile, but none was successful. In fact, when considering only the vertical profile, a two-component model cannot be made to match the data, as during the modeling process, the components tend to converge to a single scale height, resulting in a one-component model. 

\par
     However, since the existence of a halo is of primary concern, we have made substantial attempts to optimize this model.  Since the vertical profile and zeroth-moment map provide no useful constraints for the relative scale heights of the two-components or for the percentage of flux contained within the halo, we use illustrative fixed values, with the scale height of the extended component twice that of the thin component, and with 25 percent of the flux residing in the extended component. These values are chosen because they represent a reasonable scenario compared to other galaxies with halos \citep{2007AJ....134.1019O}. 
   
\par
     Through optimization of this two-component model, it is found that a reasonable fit to the data is impossible without including a warp component along the line of sight. A halo will only add to the complexity of the model and if a warped model without a halo is sufficient (or in this case, slightly better), then the simpler model should be used.  Thus, we reject the two-component model with or without such a warp.  We do however, retain a two-component model with a warp component along the line of sight (``2CW") in figures throughout this paper for illustrative purposes only.

\subsubsection{Improvement With the Addition of a Lag} \label{Section4.2.5}

\par
     At this point, prior to the addition of a lag, we place our confidence in the single component LOS warp model.  However, as may be seen in Figures~\ref{fig6} and~\ref{fig7}, the optimized models with no lag do not exactly match the data.  Adding a lag at this stage will best illustrate its contribution to the models.  We continue to include the one-component model, the flaring model with a warp component along the line of sight, as well as the two-component model with a warp component along the line of sight for illustrative purposes as well as to demonstrate the reliability of the result concerning the lag.
      
\par
    The columns in Figure~\ref{fig6} denoted with ``L's" in the titles show the models as before, but with optimal lags for each model.  Upon inspection of these plots, it becomes evident that the addition of a lag is an almost universal improvement. By adding a lag, flux in the models is drawn to the systemic side, thus duplicating the aforementioned ``T" and ``V" shapes in the data.  Furthermore, the edge of the systemic side becomes flatter in almost all of the panels as in the data, whereas in the no lag models, only the models including a line of sight warp component appear relatively flat.  Additionally, the channel maps are improved.  They narrow in the z direction at velocities further from systemic, more closely matching the data.

\par     
      The warped model with the addition of a lag of $-9^{+3}_{-2}$ km s$^{-1}$ kpc$^{-1}$ and $-9\pm$2 km s$^{-1}$ kpc$^{-1}$ in the approaching and receding haves respectively is a reasonable fit.  Error bars are estimated via visual inspection of the range of lags which could potentially match the data and under the (inexact) assumption that the models have no additional uncertainties (e. g. uncertainty in column density or rotational velocity), as making such estimates would be unwieldy. 

\par
     For the two-component warp model, a lagging halo is simulated by adding a lag in only the thick component.  The addition of the lag substantially improves the model for the receding half with a lag as steep as $-9$ km s$^{-1}$ kpc$^{-1}$, but fails to adequately duplicate the ``V" shape in Figure~\ref{fig6} in the approaching half. Even with a lag of as steep as $-$28 km s$^{-1}$ kpc$^{-1}$ this shape cannot be duplicated. Given this difficulty, a lag of $-9$ km s$^{-1}$ kpc$^{-1}$ is used in the final model to optimize the fit for the overall shape of the data.

\par
    The improvement due to the lag is present in all models.  This attests to the robustness of the lag in that, regardless of the optimal morphological model, there are features which cannot be re-created without it.  Once again, the one-component warp model with the addition of a lag appears to be the best match to the data.

\subsubsection{The Radial Variation of the Lag}\label{Section4.2.6}

\par
\begin{figure*}
\begin{center}
\figurenum{10}
\includegraphics[width=160mm]{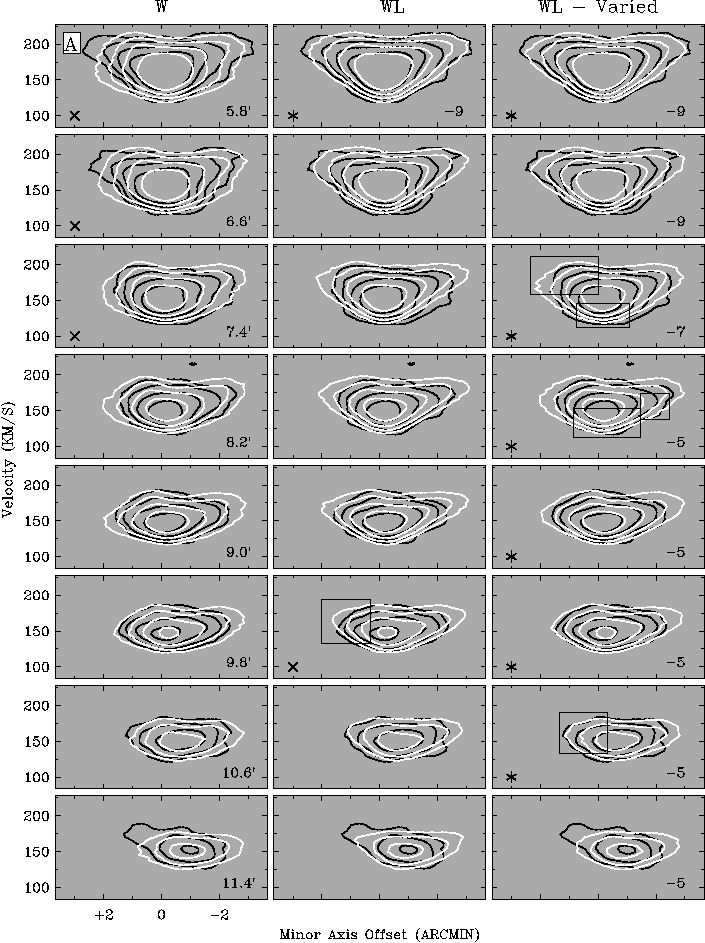}         
\caption{bv diagrams showing the warp model with no lag (left-most column), a best estimate for a constant lag (center), and a best estimate for a radially varying lag (right).  The cube represented here has a resolution of 30".  Contours begin at 3$\sigma$ (1.1 mJy bm$^{-1}$) and increase by factors of 3.  Data contours are black and model contours are superimposed in white. The lags in km s$^{-1}$ kpc$^{-1}$ for each panel of the varied warp model are given, while the lag for the constant warp model is given in the top panel only.  The offset of the slices from the center in arcminutes is given in the left-most panels and correspond to the dashed lines in Figure{~\ref{fig2}}.  The quality of fit of individual panels to the data is indicated by the convention initiated in Figure~\ref{fig6}.  Boxes are placed to show the more subtle differences between panels.  Differences which are more readily apparent to the eye are not explicitly marked in this way.  The primary issues to note are the slope of the model contours compared with those of the data on the terminal side as well as any shift up or down between them.  Also considered is the matching of the model and data contours on the high-z edges on the systemic side.  The approaching half is shown in (A) while the receding half is shown in (B).\label{fig10}}
\end{center}
\end{figure*}

    Upon establishing the existence of a lag, it must be characterized by not only the magnitude of the gradient in the z direction but also its radial variation. Using the one-component warped model, bv diagrams further from the center of the galaxy than those shown in Figure~\ref{fig6} are examined in Figure~\ref{fig10} using a 30$\times$30" cube, with the range of slice locations corresponding to the dashed lines in Figure~\ref{fig2}. The data, at these large radii, are best fit with a shallowing lag, decreasing in magnitude to $-5\pm$2 km s$^{-1}$ kpc$^{-1}$ and $-4\pm$2 km s$^{-1}$ kpc$^{-1}$ near a radius of 8' (10 kpc) in the approaching and receding halves respectively. This is most easily seen on the terminal side of each panel:  the data contours become more rounded, whereas the model with a constant lag retains too much of a ``V" shape, with data contours extending further on the terminal side than those of the model.  However, on the negative offset side of the receding half, the shallowing of the lag may be even more substantial.  This is best seen in panels corresponding to -8.2' and -9'. Uncertainties are still quantified based on the positive offset side, and $-4$ km s$^{-1}$ kpc$^{-1}$ is only given as an upper limit for the side with negative offset.  There is no strong indication of any radial variation at radii smaller than approximately 6'  (7.5 kpc). It should be noted that the S/N of the data is considerably diminished at and beyond a radius of 10' and determining the lag this far from the center approaches the limits of our modeling capabilities.  The error estimates are once again by eye.

\begin{figure*}
\begin{center}
\includegraphics[width=160mm]{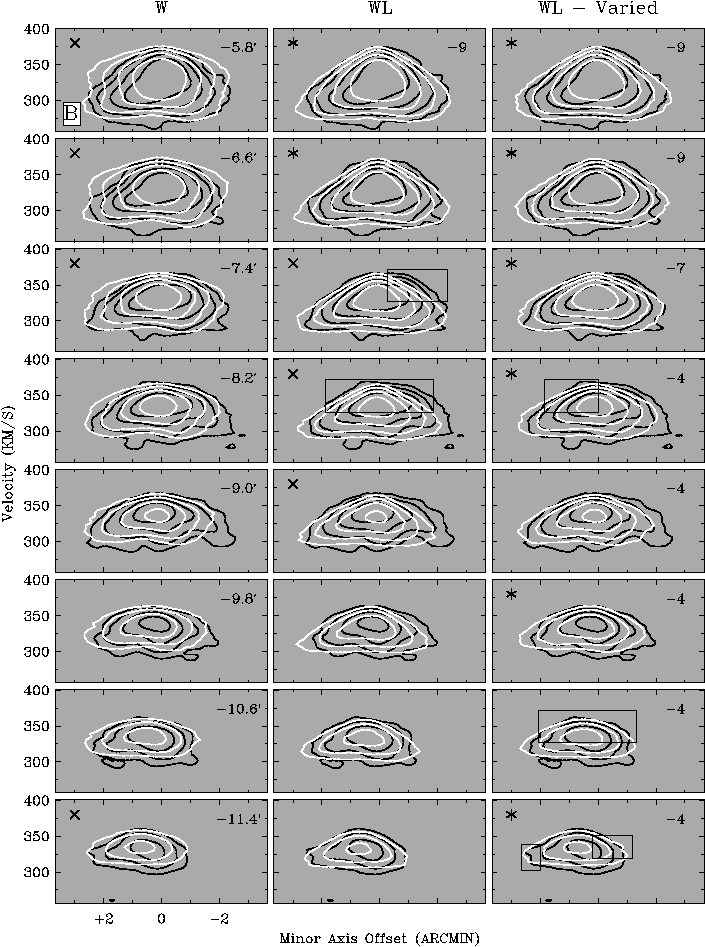}     
\label{fig10b}
\end{center}
\end{figure*}

\section{Distinct Features} \label{Section5}
\par
\begin{figure}
\begin{center}
\figurenum{11}
\includegraphics[scale=.85]{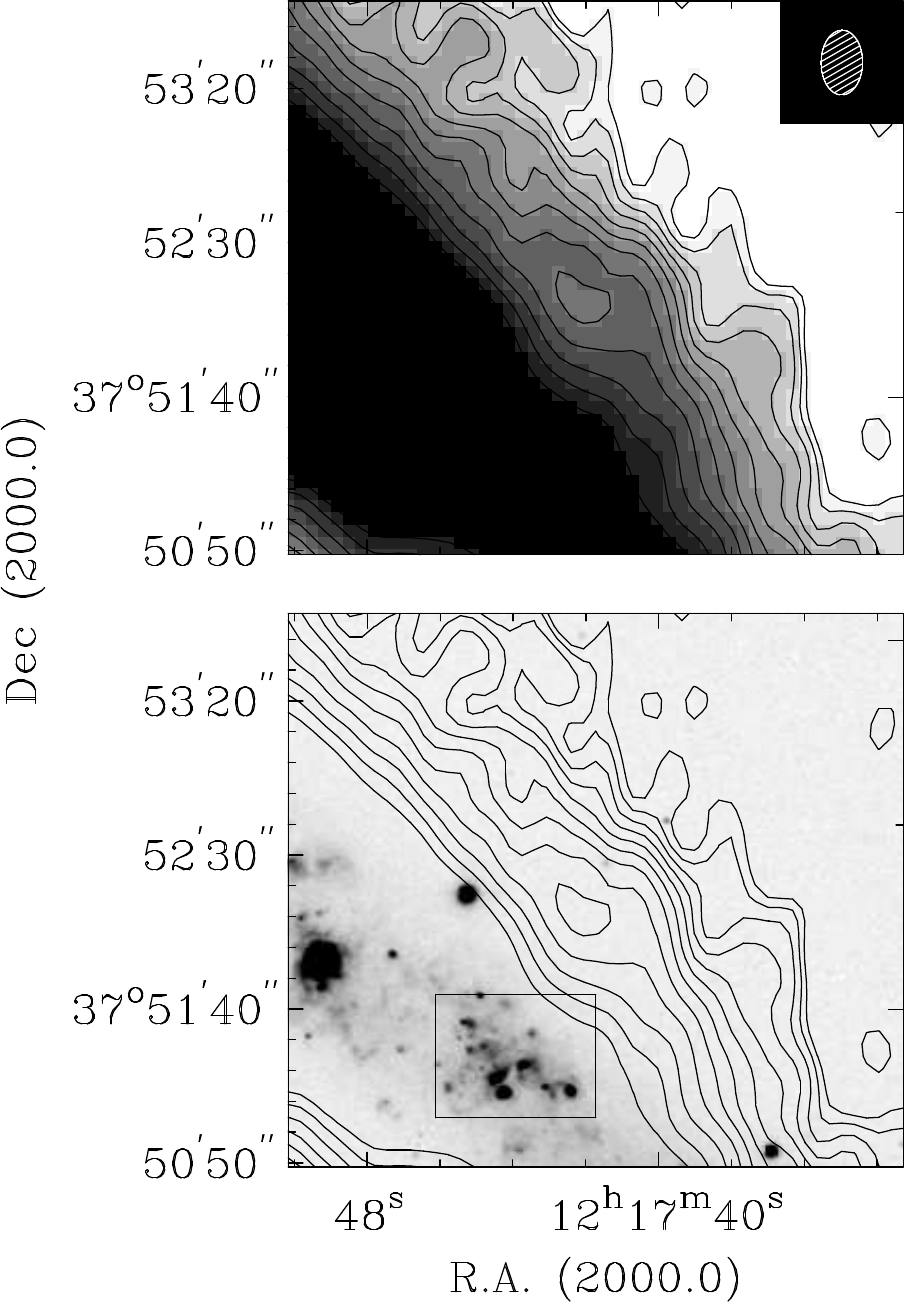}            
  \caption{A zeroth moment map of the shell-like feature on the approaching half centered at 12h17m42s, 37d53m9s and described in the text (top) and plotted over the H$\alpha$ image (bottom) with the {\sc H\,i} beam displayed in the upper right-hand corner of the top plot.  The {\sc H\,i} contours start at $4.7\times10^{18}\mathrm{cm^{-2}}$ and increase by factors of $\sqrt{2}$.  The velocity range included is 162-174 km s$^{-1}$ The box in the lower panel outlines the star forming region described throughout the text.  \label{fig11}}
\end{center}
\end{figure}
\begin{figure}
\begin{center}
\figurenum{12}
\includegraphics[scale=.55]{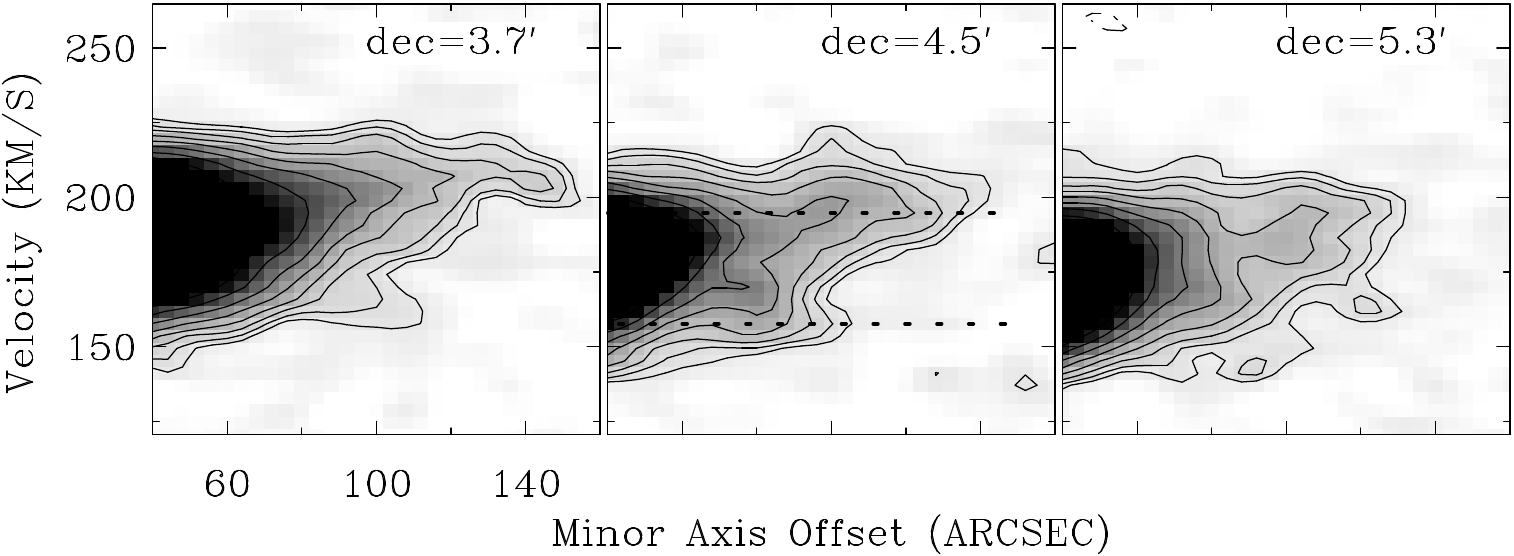}       
\caption{bv plots at the location of the shell-like feature displayed in Figure~\ref{fig11}.  The dashed lines in the center panel show line splitting at the center of the presumed shell, while the left and right panels correspond to the approximate locations of the shell walls.  Contours begin at 0.56 mJy bm$^{-1}$ and increase by factors of $\sqrt{1.8}$.\label{fig12}}
\end{center}
\end{figure}

    Aside from the global characterization of {\sc H\,i} kinematics and distribution, it is a goal of HALOGAS to characterize localized features which may be accretion events or indications of disk-halo cycling.  Furthermore, there exist prominent features in the data we wish to demonstrate are not global in the sense that our models do not attempt to reproduce them.  Perhaps the most significant is a shell-like feature located in the western quadrant of the approaching half.  An arcing feature is seen in a zeroth-moment map made by including channels ranging from 149 km s$^{-1}$ to 175 km s$^{-1}$, extending from roughly 3.6' to 5.1' along the major axis (Figure~\ref{fig11}). Corresponding extended emission is also observed as line splitting in the bv plots (Figure~\ref{fig12}), which will be discussed in $\S$~\ref{Section6.4}.

\par
\begin{figure}
\figurenum{13}
\includegraphics[width=80mm]{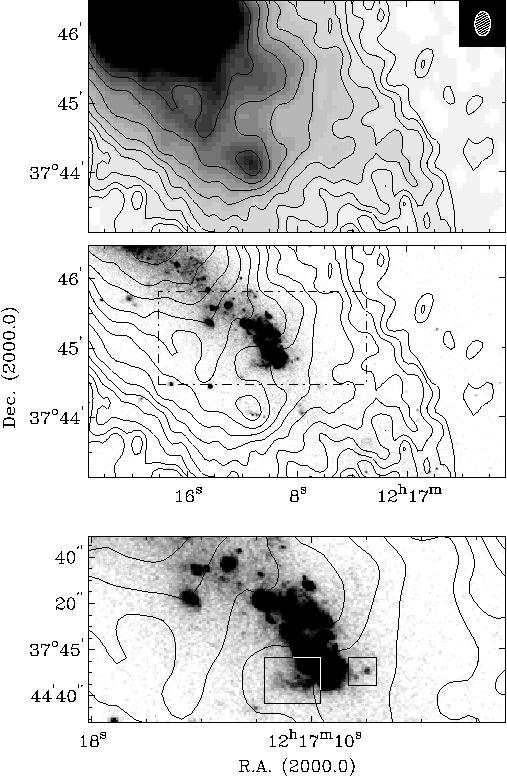}         
  \caption{A zeroth moment map of the curved feature on the receding half described in the text (top), plotted over the H$\alpha$ image (middle) as well as a zoomed-in image (bottom) of the boxed area in the H$\alpha$ image with an enhanced grayscale showing potential outflow from the {\sc H\,ii} region (in the boxed areas), which falls in a gap in the {\sc H\,i}.  The {\sc H\,i} contours start at $3.7\times10^{19}\mathrm{cm^{-2}}$ and increase by factors of $\sqrt{2}$.  The velocity range included is 281-302 km s$^{-1}$. \label{fig13}}
\end{figure}
\begin{figure}
\includegraphics[angle=270, width=80mm]{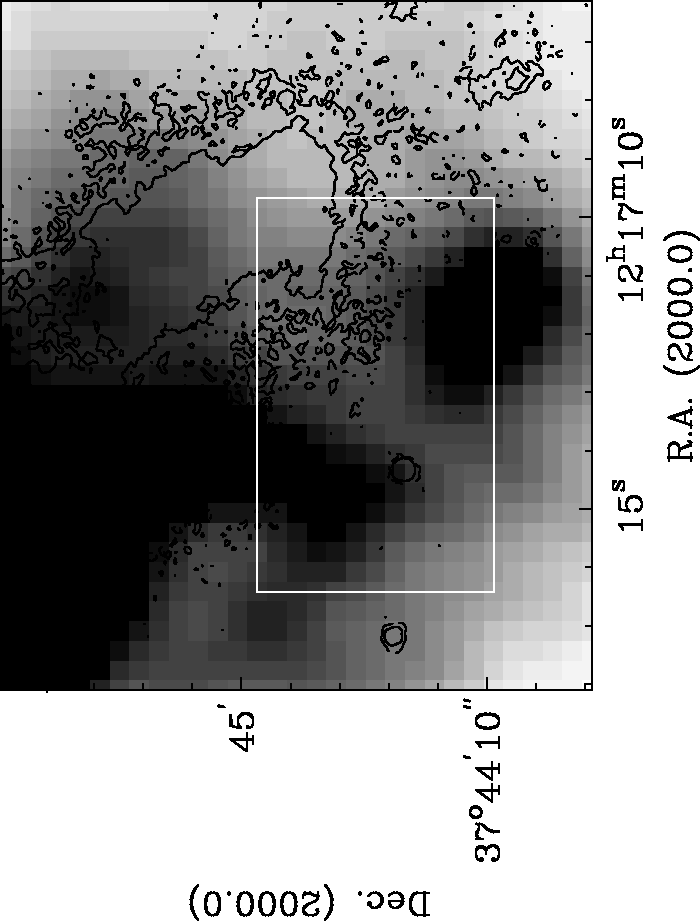}  
  \caption{An {\sc H\,i} channel map at 298 km s$^{-1}$.  3 and 7$\sigma$ H$\alpha$ contours are overlaid on an {\sc H\,i} grayscale to show the outline of the faint H$\alpha$ emission.  Notice the gap in the {\sc H\,i} near the location of the potential H$\alpha$ outflow also shown in Figure~\ref{fig13}. \label{fig14}}
\end{figure}

     An additional noteworthy feature is observed in the receding half (Figure~\ref{fig13}), roughly 4.3' to 6.8' radially from the center, and present in channels corresponding to 281-302 km s$^{-1}$.  This feature displays an elongated path curving away from the major axis in the {\sc H\,i} contours, starting from the midplane and extending beyond 1.5' above the midplane.  Furthermore, it appears that there may be a hole or vent in the {\sc H\,i} at this location, which may correlate with H$\alpha$ features (Figure~\ref{fig14}). Details of this feature will be discussed in $\S$~\ref{Section6.4}.

\par
    A narrow, pronged, faint streak may be seen from 302 km s$^{-1}$ to 355 km s$^{-1}$ in the channel maps of the receding half (Figure~\ref{fig3}).  Beyond 355 km s$^{-1}$ it becomes difficult to discern the streak from the regular emission of the galaxy which can be modeled.  However, by comparing the {\sc H\,i} along the mid-plane in these channels for each half, it is seen that the receding half extends only slightly further than the approaching.  Thus, it is possible that, rather than there existing a streak of additional {\sc H\,i}, there may instead be a dearth of {\sc H\,i} at moderately high z in these channels.  This feature will not be discussed further as we have no clear explanation.

\par
     Several small-scale shell-like features are also seen in both halves of NGC 4244, but are not shown or discussed further as they do not have substantial effects on the modeling and appear to have no immediate connection to features in the plane of the disk.

\section{Discussion} \label{Section6}

\subsection{The Warp in NGC 4244} \label{Section6.1}
\par
     Our models show that a warp starting around a radius of 12 kpc exists in NGC 4244. This warp is not only seen in the zeroth-moment map, forming a classic shallow integral shape, but also along the line of sight through our more in-depth modeling. These two-components of a warp, while representing tilts around axes off-set by 90$\,^{\circ}$ from each other, are similar in amplitude (8$\,^{\circ}$ and 4-5$\,^{\circ}$) and starting point (Figure~\ref{fig8}).  Considering both contributions, the angle included by the spin vector of the inner, flat disk and the spin vector at the outermost radius, 17.5 kpc, for which we can reliably determine inclination and position angle (i. e. the warp amplitude) is 9-9.5$\,^{\circ}$.  The starting point, being near the optical radius  is consistent with the first two rules of commonly observed warp behavior \citep{1990ApJ...352...15B}.  This agreement between two independently modeled quantities as well as the consistency with past observations attests to the reliability of the models.  

\subsection{Impact on Lag Trends} \label{Section6.2}

\par
    In addition to the presence of extra-planar gas, trends concerning its properties such as the existence and magnitude of lags are of use in determining its origins.  Although we do not detect a vertically extended {\sc H\,i} halo in NGC 4244, we do detect a lag in a thickened {\sc H\,i} layer having a scale height of approximately 550-575 pc.  This indicates that substantial extra-planar gas need not be present in order for a lag to be observed, as is potentially seen in UGC 1281 \citep{2011arXiv1103.0699K}, although for that galaxy, a more substantial line of sight warp component is favored in place of a lag.

\par
     Any correlations with DIG layers and their kinematics are also critical.  Results from \citet{2004A&A...424..485F} show agreement between the kinematics of the {\sc H\,i} and DIG layers off the plane in NGC 2403, providing an early indication that they may be connected.  \citet{2007ApJ...663..933H} discovered a trend [albeit for only 3 galaxies: NGC 891 ($-$17.5 km s$^{-1}$ kpc$^{-1}$), NGC 4302 ($-$30 km s$^{-1}$ kpc$^{-1}$) and NGC 5775 ($-$8 km s$^{-1}$ kpc$^{-1}$)] of a shallower lag with increasing star formation activity for extraplanar DIG.  At this time, H$\alpha$ data for NGC 4244 is being analyzed by additional HALOGAS members to determine the DIG kinematics and results are forthcoming (Wu et al. 2011, \textit{in prep}). Although we cannot say for sure, the low star formation rate of 0.12 M$_{\odot}$ yr$^{-1}$ in NGC 4244, which is substantially lower than those in the galaxies previously listed, indicates that if a lag in the DIG exists, then it should be steep compared to the trending galaxies.   However, the lag in {\sc H\,i} in NGC 4244 is shallower than the DIG lags for NGC 4302 and NGC 891 and slightly steeper than that found in NGC 5775.  Given this, either the trend of low star formation with steeper DIG lags will be broken if the {\sc H\,i} and DIG lags match, or the magnitudes will differ, possibly indicating different origins.  

\par
     \citet{2007ApJ...663..933H} also found that, although the lags differed substantially in the three DIG layers studied, they are in better agreement with each other when expressed in terms of km s$^{-1}$ per DIG scale height with an average value of $-$20 km s$^{-1}$ per scale height. If one considers the NGC 4244 lags in such terms, the {\sc H\,i} lag found in NGC 4244 is dissimilar to those found in the DIG layers of previously modeled galaxies.  Assuming the warp model, with scale heights of 550 pc and 575 pc in the approaching and receding halves respectively, the lag is roughly $-$5 km s$^{-1}$ per scale height. 

\par
     However, it is useful to note that possibly differing magnitudes in the {\sc H\,i} and DIG lags do not necessarily mean that they do in fact have different origins.  The {\sc H\,i} and DIG have substantially different distributions, with the {\sc H\,i} having significantly greater radial extent as may be seen in Figure~\ref{fig1}.  Since the magnitude of a lag from an internal source is somewhat dependent on where the gas is initially launched, at least in ballistic models (\citealt{2002ApJ...578...98C}, \citealt{2006MNRAS.366..449F}), it is safe to say that different radial density profiles, as are commonly seen in {\sc H\,i} and DIG layers, can potentially result in differences in the lags.
     
\par
     We now compare the magnitude of the lags in NGC 4244 with lags found in the {\sc H\,i} halos of other galaxies.  Gradients in {\sc H\,i} layers have been modeled for NGC 891 [$-$15 km s$^{-1}$ kpc$^{-1}$ \citep{2007AJ....134.1019O}], UGC 7321 [$<-$25 km s$^{-1}$ kpc$^{-1}$ \citep{2003ApJ...593..721M}] and the Milky Way [$-$22 km s$^{-1}$ kpc$^{-1}$ \citep{2008ApJ...679.1288L}, $-$15 km s$^{-1}$ kpc$^{-1}$ \citep{2010arXiv1010.3563M}].  The lag measured for NGC 4244 is substantially shallower than these.  The star formation rate in NGC 4244 of 0.12 M$_{\odot}$ yr$^{-1}$ is substantially lower than that of NGC 891 [3.8 M$_{\odot}$ yr$^{-1}$ \citep{2007AJ....134.1019O}] and the Milky Way [0.68-1.45 M$_{\odot}$ yr$^{-1}$ \citep{2010ApJ...710L..11R}], but comparable to UGC 7321 [0.15 M$_{\odot}$ yr$^{-1}$ \citep{2011A&A...526A.118H}].  Given the small number of galaxies mentioned above and the lack of any pattern concerning star formation rate and the magnitude of a lag in extraplanar {\sc H\,i}, it is still too early to establish a significant trend.

\subsubsection{The Relation Between the Lag and Star Formation Within NGC 4244}\label{Section6.2.1}

\par
    Although the lag in the {\sc H\,i} layer of NGC 4244 does not fit the trend with star formation rate seen by \citet{2007ApJ...663..933H} for DIG layers, we can ask whether such a trend is apparent {\it within} NGC 4244:  the lag is comparable on the two sides of the galaxy, but is this also true of the star formation rate?  If the \citet{2007ApJ...663..933H} trend continues, then a large contrast in star formation rate would be unexpected. We consider the star formation in each half separately by using H$\alpha$ \citep{1999ApJ...522..669H} and 24 $\mu$m MIPS data (NASA/IPAC Infrared Science Archive).  Using these, we apply the method described in \citet{2009ApJ...703.1672K} to correct for extinction.   We estimate the uncorrected H$\alpha$ flux in each half by measuring the percentage of the total H$\alpha$ flux in each and then using the total H$\alpha$ luminosity for NGC 4244 found in \citet{2008ApJS..178..247K} of $10^{40}$ erg s$^{-1}$.  Accounting for the slight difference in distances used, these estimates are $5.4\times10^{39}$ and $4.4\times10^{39}$ erg s$^{-1}$ in the approaching and receding halves respectively.  (This ordering will remain consistent throughout this section.) Assuming a monochromatic flux ($L={\lambda}L_{\lambda}$) for the 24, 70 and 160 $\mu$m data, the total infrared (TIR) fluxes are found to be $1.1\times10^{42}$ and $1.0\times10^{42}$ erg s$^{-1}$.  By assuming a Salpeter IMF and then applying a conversion between line fluxes and star formation rates supplied by \citet{2007ApJ...671..333K}, the star formation rate for each half may be determined:
\begin{equation}
SFR_{\rm M_{\odot} \rm{ yr}^{-1}}=9\times10^{-42}[L_{\rm H\alpha}+aL_{\rm TIR}]
\end{equation}
(where $a=0.0024$ and units are in erg s$^{-1}$).  These are found to be 0.064 and 0.054 $\textrm M_{\odot}$ yr$^{-1}$. The star formation rate in the approaching half is approximately 15$\%$ higher than that in the receding half.  However, the uncertainties in both the lag magnitude and the local star formation rates do not allow for an unambiguous conclusion regarding the trends seen by \citet{2007ApJ...663..933H}, but there is no indication of deviation from it.

\subsubsection{The Radial Variation of the Lag}\label{Section6.2.2}

\par
    Now we consider the radially outward decrease in the magnitude of the lag, which has yet to be fully explained.  This decrease begins just within the optical radius (near 7' or 9 kpc) and continues until just past the start of the warp.  Thus, it may be concluded that this is an effect independent of the warp. By comparing the span of this radial variation in lag with the distribution of star formation as traced in H$\alpha$ and 24 $\mu$m emission (Figure~\ref{fig1}),  The onset of the decreasing lag magnitude coincides with a sharp decline in star formation, again going against the tentative trend for extraplanar DIG rotation found by \citet{2007ApJ...663..933H}. However, it is premature to assume that star formation is the only process affecting the magnitude of the lag and other factors may be the cause.  Simple geometric arguments predict a shallowing of the lag at large radii.  The most basic interpretation of the geometry, without any consideration of gas dynamical effects, considers the relationship between the rotational velocity and the gravitational potential:  gas rotating at high z is further from the gravitational center than the gas at the same radius in the midplane, resulting in a slower rotational velocity.  This effect is most noticeable at smaller radii, where a certain height above the disk is a larger fraction of the radial distance from the center, resulting in a larger relative change in the gravitational potential, and thus a steeper lag. 

\par
    Additionally, a radial pressure gradient could affect the lag and its radial variation. According to \citet{2002ASPC..276..201B}, a radial pressure gradient directed inward may cause a steepening of the lag.  In contrast, an outward pressure gradient may result in a shallowing of the lag or even an increase in rotation velocity.  Such gradients could also vary with radius.  Given the uncertainty in the pressure gradients in question, as well as the potentially substantial impacts even small pressure gradients may have, it is difficult to say to what degree these will alter the magnitude of any given lag.  

\par
     Observationally, a radially varying lag is also seen in NGC 891 \citep{2007AJ....134.1019O}, where the gradient in the inner regions is $-$43 km s$^{-1}$ kpc$^{-1}$ and decreases to $-$14 km s$^{-1}$ kpc$^{-1}$ in outer radii.  This shallowing of 2.5 km s$^{-1}$ kpc$^{-2}$ is comparable to what the models indicate for NGC 4244, although the uncertainties (as judged by eye) make this a difficult comparison.  Furthermore, the shallowing of the lag in NGC 891 was seen closer to the center in both H{$\alpha$} \citep{2007A&A...468..951K} and {\sc H\,i}.  Thus, the origins of the shallowing of the lags in NGC 4244 and NGC 891 may be different.
 
\par
     Also displaying radial variation in the lag is the DIG in one quadrant of NGC 4302 \citep{2007ApJ...663..933H}, but the trend found in that galaxy was opposite from what we detect in the {\sc H\,i} in NGC 4244.  NGC 4302 displayed a substantial steepening of the lag, going from approximately $-$23 km s$^{-1}$ kpc$^{-1}$ to almost $-$60 km s$^{-1}$ kpc$^{-1}$ at radii outward of 4.25 kpc.

\par
     Ultimately, a larger sample of galaxies where this phenomenon is observed and modeled will greatly aid in determining its cause and nature.

\subsection{Impact on {\sc H\,i} Halo Trends} \label{Section6.3}

\par
    There is substantial evidence against the existence of an extended {\sc H\,i} halo in NGC 4244, most notably the failure of the addition of a second vertical component by itself or accompanied by a warp along the line of sight to improve the fit of the vertical profile (Figure~\ref{fig5}), and that the appearance of Figure~\ref{fig1} does not differ substantially from the 10$\times$ shallower Olling data. Furthermore, a second component added to the warp model yields no overall improvement to the bv plots or channel maps and adds an unnecessary degree of complexity. The significance of the lack of an {\sc H\,i} halo has yet to be determined.  Having such a small number of galaxies observed to this depth and modeled using these methods makes it impossible to extract any reliable trends.  This will be remedied in the near future via the HALOGAS survey and EVLA observations involving some HALOGAS team members.  For now, we can only compare the results for NGC 4244 with those of previously modeled galaxies.

\par
     Extended neutral halos have been detected in several galaxies listed below. The full resolution 1$\sigma$ rms noise per channel for each external galaxy is given in mJy bm$^{-1}$. An extended neutral halo is seen in the Milky Way \citep{2008ApJ...679.1288L}, with a scale height of 1.6 kpc \citep{2010arXiv1010.3563M} NGC 891 (\citealt{1997ApJ...491..140S}; 0.09 mJy bm$^{-1}$, 23.4$\times$16" resolution), with a scale height of 1.25-2.5 kpc \citep{2007AJ....134.1019O}, NGC 4559 with a maximum scale height of 4 kpc (\citealt{2005A&A...439..947B}; 0.52 mJy bm$^{-1}$, 12.2$\times$24.5" resolution), as well as UGC 7321 \citep{2003ApJ...593..721M} with a FWHM of 3.3 kpc; 0.36-0.40 mJy bm$^{-1}$, 16.2$\times$15.7" resolution \citealt{2003AJ....125.2455U}).  These are all substantially greater than our estimated scale height for NGC 4244 of 565 pc, although that is only for a thickened disk, rather than a disk-halo combination as seen in those galaxies listed above.  NGC 4559 and UGC 7321 are discussed below.  
     
\par
     NGC 4559 has a star formation rate; 0.69 M$_{\odot}$ yr$^{-1}$ \citep{2011A&A...526A.118H}, approximately 5 times that of NGC 4244.  Roughly 10$\%$ of the total {\sc H\,i} mass is contained in its halo \citep{2005A&A...439..947B}.  If {\sc H\,i} halos are due to star formation, then this would be consistent with NGC 4244 having a smaller halo, or as found in this work, none at all.     
     
\par
     UGC 7321 is also a galaxy with low star formation, with a rate of 0.15 M$_{\odot}$ yr$^{-1}$.  This rate of star formation is comparable to that of NGC 4244, and yet a {\sc H\,i} halo is detected in UGC 7321, but not NGC 4244.  At first glance, the two galaxies appear quite similar.  Both display thickened disks with warps in {\sc H\,i}, and by comparing with results for UGC 7321 in \citet{2010A&A...515A..63O}, their rotation curves indicate similar masses, and thus similar gravitational potentials.  Finally, neither has close companion galaxies.  In spite of these similarities, extended wing structures are seen in the vertical profiles shown in Figures 2 and 5 of \citet{2003ApJ...593..721M}, which are absent in analogous plots of NGC 4244 (not shown), drawing a clear and fundamental distinction between the morphologies of the two.  The presence of a halo in UGC 7321, a galaxy with a comparable star formation rate to that of NGC 4244, indicates that {\sc H\,i} halos \textit{are not} necessarily due solely to star formation.

\subsection{Analysis of Distinct features} \label{Section6.4} 

\par
    It is not the primary intent of this paper to perform an exhaustive study of small-scale features.  NGC 4244 is nearly devoid of prominent, energetic {\sc H\,i} shells as well as expanding features akin to those seen in more actively star forming galaxies.  However, a small number of notable features are detected (\S 5) and will be discussed.

\subsubsection{The Shell-Like Feature}\label{Section6.4.1}
\par
     Firstly, we consider the shell-like feature in the approaching half.  This feature is above and slightly offset radially from a region of star formation (Figure~\ref{fig11}), which extends between 3.5' and 4.6' along the major axis in the H$\alpha$ and 24 $\mu$m data.  The feature's proximity to a region of star formation as well as the lack of nearby external features indicate an internal origin.  To estimate the energy required to produce it, the number density must first be found.  For the calculations below we assume this feature is indeed a shell created by multiple supernovae in the disk. Following the approach taken by \citet{1993AJ....105.2098R}, we measure the peak flux in the limbs of the presumed shell to acquire a column density.  To obtain the number density, we then consider the path length along the line of sight ($l$) to be:
\begin{equation}
2\sqrt{d^{2}+2dr}
\end{equation}
Where $d$ is the shell wall thickness and $r$ is the inner radius of the shell.  Since the shell walls are not well resolved, we set $d$ equal to the beam resolution parallel to the disk of about 17".  From this, a value of 1280 pc is obtained for $l$, which yields a number density of 0.15 cm$^{-3}$. 

\par
     Now using Figure~\ref{fig12}, an estimate for the expansion velocity ($V_{\rm exp}$) is 25 km s$^{-1}$.  Together with the estimate for the shell's radius in parsecs ($R_{\rm shell}=750$ pc) and assuming a constant expansion, its age is approximately $2.9\times10^{7}$ years. 

\par
    We now calculate the energy required to produce such a shell from \citet{1974ApJ...188..501C} where $n_{\rm 0}$ is the {\sc H\,i} number density in cm$^{-3}$, $R_{\rm shell}$ is the radius of the shell in parsecs, and $V_{\rm exp}$ is in km s$^{-1}$ as described above:   
\begin{equation}
E_{\rm E}=5.3\times10^{43}n_{\rm 0}^{1.12}R_{\rm shell}^{3.12}V_{\rm exp}^{1.4} \textrm{ erg}
\end{equation}
 Given these estimates, we find that a single burst of energy of roughly $5.4\times10^{53}$ ergs or the equivalent of 540 supernovae would be required to produce this shell. (assuming an energy of 10$^{51}$ ergs per supernova) If the shell instead formed from a continuous supply of energy as described in \citet{1987ApJ...317..190M}:
\begin{equation}
E_{\rm C}=1.16\times10^{41}n_{\rm 0}R_{\rm shell}^{5}t_{\rm 7}^{-3} \textrm{ erg}
\end{equation}
 where $t_{\rm 7}$ is the age in units of $10^7$ years, this would only require $1.7\times10^{53}$ ergs or 170 supernovae. 

\par
     Now considering the H$\alpha$ luminosity of this region we again use the total H$\alpha$ luminosity for NGC 4244 found in \citet{1999ApJ...522..669H}.  The fraction of the total H$\alpha$ emission originating from this region is approximately 4$\%$.  Thus, its luminosity, uncorrected for extinction is found to be $4.0\times10^{38}$ erg s$^{-1}$.

\par
     To correct for extinction, we examine the 24 $\mu$m data.  Summing the flux over the same region as for the H$\alpha$ data and once again assuming a monochromatic flux, the 24 $\mu$m luminosity is approximately $4.8\times10^{39}$ erg s$^{-1}$.

\par
     Using the method described in \citet{2007ApJ...671..333K}, we determine the star formation rate, this time with $a=0.038$ and replacing $L_{\rm TIR}$ with the 24 $\mu$m luminosity to account for localized regions.
\par
     The star formation rate for this region is found to be 0.0046 $M_{\odot}yr^{-1}$, which would account for 3.8$\%$ of the ongoing star formation in NGC 4244. However, \citealt{2007ApJ...671..333K} caution that this conversion is intended for entire galaxies rather than localized regions, which limits its physical meaning.

\par
     Using this star formation rate (assuming it is constant) as well as the estimated age for the shell, a calculation akin to that presented in \citet{1997ApJ...476..144M} yields an estimated 650-700 supernovae to have gone off in the region, rendering supernovae a very plausible source.

\subsubsection{Curved Feature}\label{Section6.4.2}

\par
     Attending now to the curved feature in the receding half shown in Figure~\ref{fig13}, we examine its potential connection to star formation.   It is likely that there is some correlation between this feature and the underlying star forming region seen in the both H$\alpha$ and 24 $\mu$m data.  The best resolved emission is seen in the H$\alpha$ data, which shows a curious extension of emission from possibly blown out gas (Figure~\ref{fig13} bottom panel, highlighted in boxes) which to some extent fills in the gap defined by the {\sc H\,i} contours, indicating that the {\sc H\,i} gas may have been forced outward in that direction.  This structure does not appear in the 24 $\mu$m data (not shown).  When examining individual channel maps, it can be noted most prominently at 294 km s$^{-1}$ and 298 km s$^{-1}$, a gap in the {\sc H\,i} forms, once again suggesting a channeling of the gas away from the midplane and into higher z in this area (Figure~\ref{fig14}). This hole in the {\sc H\,i} emission may be analogous to those seen in other galaxies such as NGC 4559 \citep{2005A&A...439..947B}. However, it is possible that the features are coincidental.

\subsection{Comparison to Previous NGC 4244 Models} \label{Section6.5} 
\par
     Our results concerning NGC 4244, using direct visual comparisons between models and data, represent an improvement over those obtained using the methods found in \citet{1996AJ....112..457O}.  Firstly, the warp resulting from models presented in this paper is more pronounced than that previously indicated.  Furthermore, the inclination in central parts of the galaxy is higher (88$\,^{\circ}$ as opposed to 84.5$\,^{\circ}$). The vertical profiles in Figure~\ref{fig5} show a comparison between the models.  It should be noted that in this case, the lower inclination in the Olling model does not create the expected rounding near the top of the vertical profile and there is a lack of {\sc H\,i} in the wings.  However, this is because the {\sc H\,i} distribution near the center of the Olling model is much higher than in the models presented here.  This is also seen in bv diagrams of the Olling model (not shown). Additionally, there is no need for a flaring layer in the current models in order to fit the data, although one may be present.  These discrepancies cannot be accounted for by improved observations alone and appear to result from differences in technique.  

\section{Summary and Conclusions} \label{Section7}

\par
    We present tilted ring models based on deep 21-cm observations of NGC 4244.  In these observations, we note a thickened {\sc H\,i} disk, with thinning at outer radii and evidence for a warp.  We also note asymmetries in the approaching and receding halves as well as localized features.  The models tested, in order of approximate increasing complexity, include a single disk with constant inclination, a single disk with a warp component along the line of sight, a single flaring disk, a thin and thick disk, and combinations of these.

\par
     From these models, we conclude that the disk of NGC 4244 is substantially warped both perpendicular to as well as along the line of sight.  We do not detect an extended {\sc H\,i} halo, but with a scale height of 0.6 kpc, the disk is quite thick.  If {\sc H\,i} halos are due to disk-halo flows, then given the low star formation rate in NGC 4244 compared to other galaxies with {\sc H\,i} halos, the lack of a halo is unsurprising.  However, it is still to early to establish a trend, which will hopefully become apparent when additional cases are considered.

\par     
     In spite of not detecting an extended halo, we do detect a vertical lag in rotation speed of $-9^{+3}_{-2}$ km s$^{-1}$ kpc$^{-1}$ and $-9\pm$2 km s$^{-1}$ kpc$^{-1}$ in the approaching and receding halves respectively. The magnitude of this lag decreases outward with radius to $-5\pm 1$ km s$^{-1}$ kpc$^{-1}$ (approaching) and $-4\pm 2$ km s$^{-1}$ kpc$^{-1}$ (receding) near 10 kpc, although subtle changes are within the error estimates.  Beyond 13 kpc, the S/N ratio no longer allows for a reliable lag estimate, so we cannot discern whether the lag vanishes entirely at some point.  

\par
     This is potentially the first instance where models favor a lag in {\sc H\,i} in a galaxy where an {\sc H\,i} halo does not exist.   There may be a lag in UGC 1281 \citep{2011arXiv1103.0699K}, in which a halo is also absent, but in that case the lag cannot be distinguished from a warp component along the line of sight without deeper observations.  Given fluctuations in the position angle seen in the zeroth and first moment maps, it is possible that the warp in NGC 4244 begins closer to the center than in the models presented, which may indeed lessen or eliminate the need for a lag.  However, such a scenario is unlikely as these fluctuations may instead be due to localized features, a bar or spiral arms, which would be more in-line with current understanding of warps (\citealt{1990ApJ...352...15B}, \citealt{2011arXiv1101.1771V}).

\par
     Modeling aside, we observe two substantial localized features which may correlate with star formation:  a shell-like feature and an unexplained curved structure originating in the midplane and possibly connected with a feature extending up to a height of 1.5 kpc.  A third distinct feature, a faint {\sc H\,i} streak is anomalous.  Finally, no additional substantial energetic features are detected, although numerous small-scale shell-like features are seen, but not displayed.

\par
     We hope to use information gathered from our models of NGC 4244 in addition to information from the rest of the HALOGAS galaxies to further the development of trends concerning {\sc H\,i} kinematics, warps, extra-planar gas and any connections with disk-halo interactions, as well as accretion.

\section{Acknowledgments} \label{Section8}

\par 
     We would like to thank the operators at WSRT for overseeing the observations.  The Westerbork Synthesis Radio Telescope is operated by the ASTRON (Netherlands Institute for Radio Astronomy) with support from the Netherlands Foundation for Scientific Research (NWO).  This research has made use of the NASA/IPAC Infrared Science Archive, which is operated by the Jet Propulsion Laboratory, California Institute of Technology, under contract with the National Aeronautics and Space Administration. We acknowledge the HALOGAS team, especially Filippo Fraternali, Renzo Sancisi, Gyula  J\'{o}zsa, Paolo Serra, Tom Oosterloo, and Robert Benjamin for help and comments during the modeling process as well as Rene Walterbos for providing the H$\alpha$ image used in this analysis.  Finally, we thank the anonymous referee for useful comments and suggestions.  This material is based on work partially supported by the National Science Foundation under grant AST-0908106 to R.J.R.  G.G. is a postdoctoral researcher of the FWO-Vlaanderen (Belgium).  P.K. is supported by the Alexander von Humbolt Foundation.

\bibliographystyle{apj}
\bibliography{NGC4244.Zschaechner.arXiv.bib}

\end{document}